\documentclass[useAMS,usenatbib]{mn2e}

\usepackage[dvips]{graphicx}
\usepackage{amssymb}
\usepackage{amsmath}
\usepackage{amstext}

\newcommand{\CHb}{$c$(H$\beta$)}
\newcommand{\Wabs}{$W_{abs}$}
\newcommand{\Te}{$T_{e}$}
\newcommand{\Ne}{$n_{e}$}

\newcommand\ion[2]{#1~{\sc {#2}}\relax}        
\newcommand\ioni[2]{${\rm #1^{#2}}$}           

\newcommand{\cmc}{{\rm cm$^{-3}$}}
\newcommand{\kms}{{\rm km~s$^{-1}$}}

\newcommand{\abox}{12+log(O/H)}
\newcommand{\lno}{log(N/O)}
\newcommand{\nodata}{...}

\newcommand{\HII}{H~{\sc ii}}
\newcommand{\HI}{H~{\sc i}}

\newcommand{\Ha}{H$\alpha$}
\newcommand{\Hb}{H$\beta$}
\newcommand{\Hg}{H$\gamma$}
\newcommand{\Hd}{H$\delta$}
\newcommand{\WHa}{$W_{\rm H\alpha}$}
\newcommand{\WHb}{$W_{\rm H\beta}$}
\newcommand{\WHg}{$W_{\rm H\gamma}$}
\newcommand{\WHd}{$W_{\rm H\delta}$}

\newcommand{\Mo}{$M_{\odot}$}
\newcommand{\Moy}{$M_{\odot}$\,yr$^{-1}$}
\newcommand{\MHII}{$M_{\rm H\,II}$}

\newcommand{\mnras}{MNRAS}
\newcommand{\apj}{ApJ}

\title[A possible localized chemical pollution by Wolf-Rayet stars in IC~10]{The ionized gas at the center of IC~10: 
A possible \\ localized chemical pollution by Wolf-Rayet stars\thanks{Based on observations collected at the Centro Astr\'onomico 
			       Hispano Alem\'an (CAHA) at Calar Alto, operated jointly by the 
			       Max-Plank Institut f\"ur Astronomie and the Instituto de 
			       Astrof{\'{\i}}sica de Andaluc{\'{\i}}a (CSIC).}}
\author[\'A.R. L\'opez-S\'anchez et al.]
       {\'A.R. L\'opez-S\'anchez$^1$\thanks{Visiting Astronomer at the Instituto de Astrof\'{\i}sica de Canarias. E-mail: Angel.Lopez-Sanchez@csiro.au}, A. Mesa-Delgado$^{2,3}$,  
        L. L\'opez-Mart{\'{\i}}n$^{2,3}$ and C. Esteban$^{2,3}$\\ 
	$^1$CSIRO Astronomy and Space Science / Australia Telescope National Facility, PO BOX 76, Epping, NSW 1710, Australia\\
	$^2$Instituto de Astrof\'\i sica de Canarias, E-38200 La Laguna, Tenerife, Spain \\
	$^3$Departamento de Astrof\'{\i}sica, Universidad de La Laguna, E-38205 La Laguna, Tenerife, Spain
		}

\begin{document}

\date{Accepted: 2010 10 08. Received: 2010 09 19; in original form 2010 10 08}
\pagerange{\pageref{firstpage}--\pageref{lastpage}} \pubyear{2010}

\maketitle
\label{firstpage}

\begin{abstract}
 
We present results from integral field spectroscopy with the Potsdam Multi-Aperture Spectrograph at the 3.5m telescope at Calar Alto Observatory  of the intense star-forming region [HL90]~111 at the center of the starburst galaxy IC~10. We have obtained maps  with a spatial sampling of  $1\arcsec\times1\arcsec$ = 3.9~pc~$\times$~3.9~pc of \mbox{different} emission lines and analyzed the extinction, physical conditions, nature of the ionization, and chemical abundances of the ionized gas, as well determined locally the age of the most recent star-formation event. By defining several apertures, we study the main integrated properties of some regions within [HL90]~111. Two contiguous spaxels show an unambiguous detection of the broad \ion{He}{ii}~$\lambda$4686 emission line, this feature seems to be produced by a single WNL star. We also report a probable N and He enrichment in the precise spaxels where the WR features are detected. The enrichment pattern is roughly consistent with that expected for the pollution of the ejecta of a single or a very small number of WR stars. Furthermore, this chemical pollution is very localized ($\sim2\arcsec\sim7.8$~pc) and it should be difficult to detect in star-forming galaxies beyond the Local Volume. We also discuss the use of the most-common empirical calibrations to estimate the oxygen abundances of the ionized gas in nearby galaxies from 2D spectroscopic data. The ionization degree of the gas plays an important role when applying these empirical methods, as they tend to give lower oxygen abundances with increasing ionization degree.  
\end{abstract}



\section{Introduction} \label{intro}

Dwarf and irregular galaxies are ideal laboratories to study the interaction between massive stars and the interstellar medium (ISM). Indeed, massive stars do not only provide the energetic radiation that ionizes the \HI\ clouds into \HII\ regions, but they also inject mechanical energy into the surrounding medium, both via strong stellar winds and supernova explosions, which create shells and bubbles and may induce the birth of new stars. The effects of the massive stars in the ISM are especially strong when they are gathered together in large clusters; their actions can even affect the nearby inter-galactic medium (IGM). Strong star-forming regions, which may be found in galaxies covering a wide range of masses, may enrich both the ISM and the IGM. Actually, it is still unclear whether the chemical enrichment provided by massive stars is an instantaneous event or if the new synthesized heavy elements must cool before becoming part of the ISM \citep*[e.g.,][]{TenorioTagle96,KS96,KS97,KS98, Kobulnicky97,Henry00,vZH06,LSE10b}. All these processes have a large effect on the formation and evolution of galaxies.

\bigskip

\begin{figure*}
\centering
   \includegraphics[width=1\linewidth]{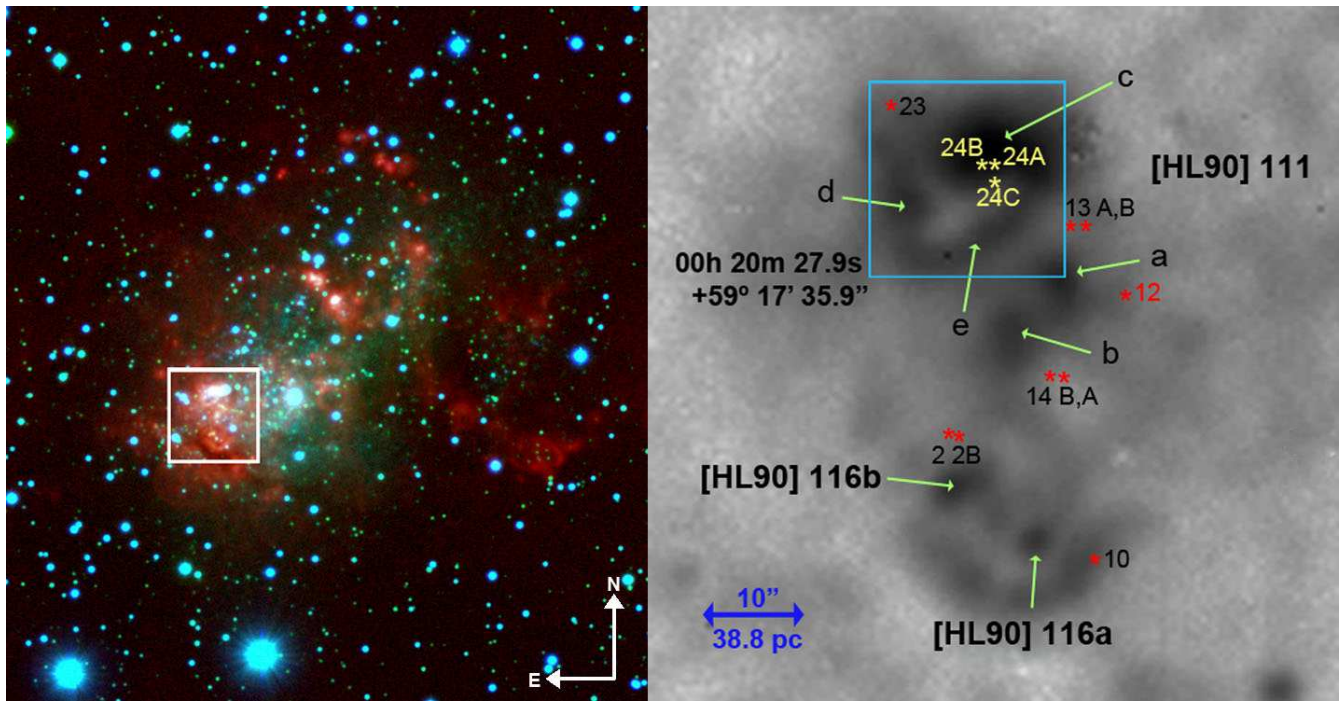}
   \caption{\label{mapa} (\emph{Left}) False-color image of IC~10 obtained combining data provided by \citet{GildePaz03} in $B$ (blue), $R$ (green) and \Ha\ (red) filters. The FoV is 6.8$\arcmin\times$7.4$\arcmin$, which covers the majority of the BCDG as it is seen in optical light. Notice the long and extensive ionized gas emission throughout all the galaxy. The main star-forming region is shown by a white box.  (\emph{Right})  Enlargement of the main star-forming region within IC~10 showing the pure \Ha\ emission (continuum-subtracted) of the galaxy \citep{HunterElmegreen04}. Its FoV is 57$\arcsec\times$57$\arcsec$. The blue box corresponds to the field of view of the IFU used with PMAS, which has a size of 16$\arcsec\times$16$\arcsec$ and covers the intense star-forming region  [HL90]~111c \citep[][see their Fig.~2b]{HL90}. The identification of the different \HII\ regions follows the names given by these authors. The positions of the individual WR stars known in IC~10 \citep{Crowther03} are shown by stars. Three of them, [MAC92] 24~A, B and C (yellow stars), are located at the center of our observed area.}
 \end{figure*}

Nearby dwarf galaxies have the advantage that they can be observed at high spatial resolution and even, in some cases, their stars are resolved. This allows a detailed ana\-lysis of the interplay between the neutral and ionized components of the ISM and the massive stars \citep[e.g.,][]{ThurowWilcots05, vanEymeren09a,vanEymeren09b,vanEymeren10, Monreal-Ibero+10} and constrains the kinematical and chemical evolution of the gala\-xies. Recently, our group confirmed a localized chemical enrichment in the nearby (4~Mpc) blue compact dwarf galaxy (BCDG) NGC~5253 \citep{LSEGRPR07}. The nitrogen pollution (and probably also helium) found in a particular area of the galaxy seems to be a consequence of the ejecta of Wolf-Rayet (WR) stars. Evidences of possible nitrogen pollution has been only reported in some few starbursts, but the majority of them also showing WR features \citep{Kobulnicky97, Pustilnik04, James09, LSE10a, LSE10b}. However, NGC~5253 remains as the only known star-forming galaxy where a localized chemical enrichment has been definitively established \citep{WalshRoy89, Kobulnicky97, LSEGRPR07, Monreal-Ibero+10}.

The advent of the 2D spectroscopy allows to explore this issue more efficiently, as this technique permits to record simultaneously the optical/near-infrared (NIR) spectra of regions within the galaxies without the limitations of long-slit observations. In particular, spectroscopy of dwarf, star-forming galaxies using integral field units (IFU) provides a very powerful tool to constrain the physical, chemical and kinematical properties of the ionized gas within these objects \citep[e.g.,][]{GarciaLorenzo08, Lagos09, Cairos09a, Cairos09b, Cairos10, James09, James10}, to look for the places where massive stars are located, to study the underlying stellar population and its kinematics via the analysis of the continuum and the absorption lines, and finally to combine all the results to understand the feedback between massive stars and the ISM \citep[e.g.,][]{ThurowWilcots05, Kehrig08, AlonsoHerrero09, Bordalo09, Monreal-Ibero+10}. Here, we present an optical 2D spectroscopical ana\-lysis of a particularly interesting and intense star-forming region within the nearest starburst, the galaxy IC~10. Although our main aim is to look for a localized chemical enrichment within the \HII\ region as a consequence of the massive stars, we will also analyze the extinction, ionization structure, chemical abundances and age of the most recent star-formation burst in this particular area of IC~10.

Sometimes defined as an irregular galaxy, IC~10 (Fig.~\ref{mapa}) actually is the only known starburst within the Local Group. Because of its low mass \citep{Mateo98}, high star-formation activity \citep[i.e.,][]{HL90, Leroy06}, low metallicity --\abox=8.26, \cite{Skillman89, Garnett90, MagriniGoncalves09}--, its very extended, clumpy, and peculiar distribution of neutral gas \citep[Juette et al., in prep;][]{ShostakSkillman89, WilcotsMiller98}, high far-infrared emission \citep{MelisseIsrael94}, non-thermal radio-continuum emission \citep{YangSkillman93}, and the relative importance of its young stellar population \citep[i.e.,][]{Sakai99, Borissova00, Hunter01, Vacca+07} which includes a large number of WR stars \citep{Massey92, MasseyJohnson98, MasseyHolmes02}, IC~10 is usually classified as a BCDG \citep{Richer+01}. The high carbon-rich WR stars to nitrogen-rich WR stars ratio in IC~10 (the WC/WN ratio), which is unusually high for its metallicity \citep{Crowther03}, suggests that the galaxy has experience a brief but intense galaxy-wide burst of star-formation within the last 10~Myr. The starburst was probably triggered by gas falling in from the large, extended \HI\ reservoir \citep{Saito92, WilcotsMiller98}.

The determination of the distance to IC~10 has been challenging because of its proximity to the Galactic plane ($l=119.0^{\circ}$, $b=-3.3 ^{\circ}$). Recently, \cite{Vacca+07} and \cite{Sanna+08} used the tip of the red giant branch (TRGB) to compute a distance of 0.79 and 0.81~Mpc, respectively, to this galaxy. Throughout this paper, we will consider that IC~10 is located at 0.80~Mpc, and hence the spatial scale is 1$\arcsec$ = 3.88~pc.

The structure of this paper is the following. In \S\ref{obsred} we describe the observations and the reduction procedure.  \S\ref{medpmas} presents the measurements of the emission lines and how the reddening correction was performed. \S\ref{resul} contains the analysis of the ionized gas (extinction, physical conditions, nature of the ionization and chemical abundances) as well as the localization of the WR stars and the determination of the age of the most recent star-formation event.  \S\ref{discu} discusses the two main results found in our analysis: the detection of a probable N and He enrichment in the precise place where the WR features are detected and a warning about the use of empirical calibrations to estimate the oxygen abundances of the ionized gas in nearby galaxies from 2D spectroscopic data. We summarize our conclusions  in \S\ref{conclu}.

  \begin{figure}
\centering
 \includegraphics[width=0.88\linewidth,angle=90]{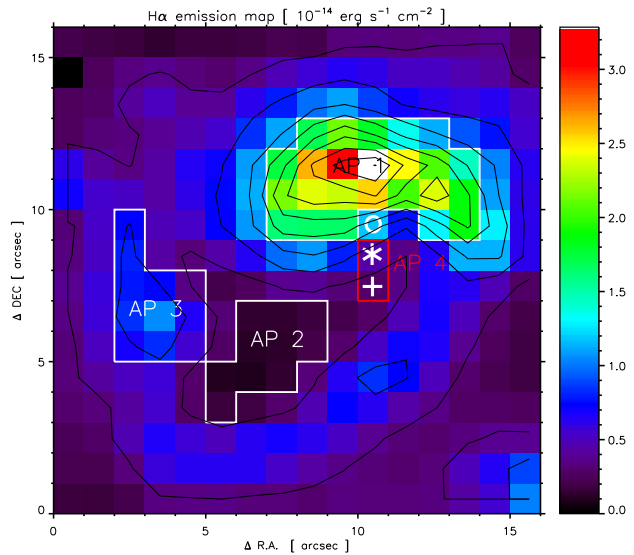}
  \caption{   \label{halpha} \label{apertures}
\Ha\ flux map of the intense star-forming region [HL90]~111 \citep{HL90} from our PMAS data, which has a size of 16$\times$16 pixels (each pixel has a size of $1\arcsec\times1\arcsec$). The flux is in units of erg~cm$^{-2}$~s$^{-1}$ and it is not corrected for extinction. The positions of the broad \ion{He}{ii}~$\lambda$4686 emission feature, which indicates the presence of WR stars, are shown by a star (strong blue WR bump) and a cross (detection of the blue WR bump). The \emph{o} symbol indicates the position of a tentative detection of the nebular \ion{He}{ii} $\lambda$4686 emission line. The four different regions used to perform a more detailed analysis of the ionized gas are also shown.}
 \end{figure}

\section{Observations and Data Reduction} \label{obsred}

The massive star-forming region [HL90]~111 \citep{HL90} , also known as [LMR79] \#1 \citep{Lequeux79}, within the starburst galaxy IC~10 was observed on 2007 October 14 at Calar Alto Observatory (Almer\'ia, Spain), using the 3.5m Telescope with the Potsdam Multi-Aperture Spectrometer \citep[PMAS,][]{rothetal2005,kelz+06}. PMAS is an integral field spectrograph, with a lens array of 16$\times$16 square elements, connected to a bundle of fiber optics, whose 256 fibers are re-arranged to form a pseudoslit in the focal plane of the spectrograph. We used the standard lens array IFU of 16$\times$16 arcsec$^2$ field of view (FoV) with a sampling of 1$\arcsec$ which, at the distant to IC~10, provides a spatial sampling of 3.88~pc/arcsec and a FoV of 62.1$\times$62.1~pc$^2$. Most of the optical range was cove\-red with the V600 grating using two grating rotator angles: $-72^{\circ}$, covering from 3500 to 5070~\AA; and $-68^{\circ}$, covering from 5700 to 7150~\AA. The effective spectral resolution was 3.6 \AA. The blue and red spectra have a total integration time of 1800 seconds in both cases. The position of the IFU covering the massive star-forming region [HL90]~111 is showed in Figure~\ref{mapa}. Its center has the coordinates $\alpha$=$00^{\rm h}\,20^{\rm m}\,27\fs9$, $\delta$=59\degr\,17\arcmin\,35.9\arcsec. The reconstruction of the \Ha\ map using our 2D PMAS spectroscopy (Fig.~\ref{halpha}) very nicely matches the \Ha\ image provided by \cite{HunterElmegreen04} and showed in the right panel of Fig~\ref{mapa}. 
The PMAS data showed two spaxels which poor light transmission, [1,15] and [8,16], which we will not consider in our analysis.

The giant \HII\ region [HL90]~111c (see Fig.~\ref{mapa}) harbors one of the brightest WR stars in the galaxy, that was named [MAC92]~24 by \cite{Massey92}. Originally identified as a WN star by \cite{Richer+01} because of the detection of the broad \ion{He}{ii}~$\lambda$4686 emission and the lack of the \ion{C}{iv} $\lambda$4650 emission, some authors suggested that this object is 
a blend of several stars \citep{MasseyHolmes02, Crowther03, Vacca+07}. In particular, \cite{Crowther03} noted that there were three closely spaced sources, [MAC92] 24A, B, and C (see Fig~\ref{mapa}), all located within $1\arcsec-2\arcsec$ at the south of [HL90]~111c. Later, \cite{Vacca+07} used ground-based laser guide star adaptive optics at the Keck~II telescope to resolve [MAC92]~24A in at least three bright, blue sources and [MAC92]~24B into two blue sources, suggesting that  there are at least four robust WN candidates in the field. This hypothesis should be confirmed using optical spectroscopy. And, indeed, our 2D PMAS spectra indicate the detection of the blue WR bump precisely at the two spaxels where the WN star candidates [MAC92]~24~B,C should be located. Their position is indicated with a star (spaxel [11,8], which shows a strong blue WR bump) and a cross (spaxel [11,9], where the WR bump is also detected but not as intense as in the previous position) in Fig~\ref{halpha}, as we discuss below. The maximum of \Ha\ emission provided by our 2D spectroscopic data is located just  $1\arcsec-2\arcsec$ at the north of the position of the WR stars, and it coincides with the center of the intense radio-continuum source reported by \citet{YangSkillman93}, [YS93] 001744.6+590101.
 
Calibration images were obtained during the night: arc lamps for the wavelength calibration and a continuum lamp needed to extract the 256 individual spectra on the CCD. Observations of the 
spectrophotometric standard stars BD~$+$28$\rm ^o$4211, Feige 110 and G~191-B2B \citep{oke90} were used for flux calibration. The absolute error of this calibration was of the order of 5\%. Night was 
photometric and the typical seeing during the observations was 1$\arcsec$. 
 
The data were reduced using the {\sc iraf}\footnote{{\sc iraf} is distributed by NOAO which is operated by AURA Inc., under cooperative agreement with NSF.} reduction package {\sc specred}. After bias subtraction, spectra were traced on the continuum lamp exposure obtained before each science exposure, and wavelength calibrated using HgNe arc lamp. The continuum lamp and sky flats were used to determine the response of the instrument for each fiber and wavelength. Finally, for the standard stars we have co-added the spectra of the central fibers and compared with the tabulated one-dimensional spectra.
 
The effect of the differential atmospheric refraction (DAR) is negligible because of both the low airmass (between 1.08 and 1.11) in which IC~10 was observed and the short exposure time ($\sim$1 hour).

\section{Line measurements} \label{medpmas}
 
\begin{figure}
   \centering
   \includegraphics[width=0.74\linewidth,angle=90]{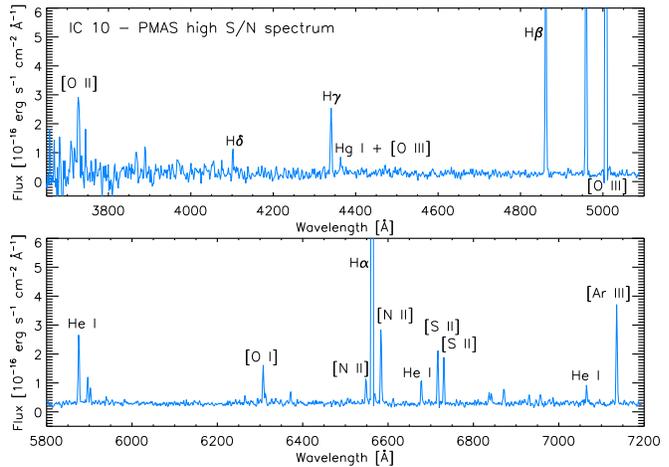}
   \caption{   \label{spectra}
   Example of the optical spectrum obtained at the center of the BCDG~IC~10 using a single fiber of our PMAS data. In particular, we plot a spectrum with a high S/N, which corresponds to the maximum of \Ha\ emission (spaxel [11,12]). The most important emission lines are labeled. 
   The spectrum has been corrected for radial velocity, but not for extinction or sky emission. In particular, note the \ion{Hg}{i} contamination at $\lambda$4359.} 
 \end{figure}

We measured all the emission lines detected in the spectra: hydrogen Balmer lines (\Ha, \Hb, \Hg, \Hd, and, in some cases, H8 and H9), helium lines (\ion{He}{i} $\lambda$4471, $\lambda$5875, $\lambda$6678 and $\lambda$7065) and the typical collisional excited lines of metallic elements: [\ion{O}{ii}]~$\lambda\lambda$3724,29 (blended), [\ion{O}{iii}]~$\lambda\lambda$4959,5007, [\ion{N}{ii}] $\lambda\lambda$6548,83, [\ion{S}{ii}]~$\lambda\lambda$6716,31, [\ion{Ar}{iii}]~$\lambda$7136 and [\ion{Ne}{iii}]~$\lambda$3869 and $\lambda$3967 (blended with H7). Figure~\ref{spectra} shows an example of the spectra provided by our PMAS observations in which all important emission lines are labelled. It corresponds to the spectrum obtained in the region with the maximum of \Ha\ emission (spaxel [11,12]).
  
Line fluxes were measured integrating all the emission between the limits of the line and over a local adjacent continuum. All these measurements were made with the {\sc splot} routine of {\sc iraf} and using our own scripts to achieve an automatic process \citep[see details in][]{mesadelgadoetal09}. However, because of the faintness of many of the emission lines, a visual inspection of the spectra was needed to get a proper estimation of the adjacent continuum and the line flux in these cases. We want to emphasize here the importance of this manual step because, although automatic procedures usually work well when measuring high signal-to-noise emission lines, they could give a wrong estimate of the flux of faint emission lines.
  
The errors associated with the line flux measurements were determined following the error criterion described in \cite{mesadelgadoetal08}. The final error of a line was the quadratic sum of the error in its flux measurement and the error in flux calibration. In order to avoid spurious weak line measurements, we imposed three criteria to discriminate between real features and noise: 1) line intensity peak over 2.5 times the sigma of the continuum; 2) FWHM($\lambda$) $>$ 1.5$\times$FWHM(\ion{H}{i}) and FWHM($\lambda$) $<$ FWHM(\ion{H}{i})/1.5; and 3) F($\lambda$) $>$ 0.001$\times F$(H$\beta$). 
  
All line fluxes of a given spectra have been corrected for extinction using the theoretical H$\alpha$/H$\beta$ ratio expected for case B recombination given by \citet{SH95} assuming the physical conditions of \Te=10000~K and \Ne=100~\cmc, the expected values for region [HL90]~111 in IC~10 (see \S\ref{condpmas}), [\Ha/\Hb]$_{\rm theo}$=2.86. We used the reddening function, $f(\lambda)$, normalized to H$\beta$ derived by \cite{Cardelli89} to correct for reddening each emission line, following the expression
 \begin{equation}
 \frac{I(\lambda)}{I({\rm H\beta})} =  \frac{F(\lambda)}{F({\rm H\beta})} \times 10^{c({\rm H\beta})\times f(\lambda)},
 \end{equation}
being $I$($\lambda$) and $F$($\lambda$) the real and the observed fluxes, respectively, of the emission line and \CHb\ the reddening coefficient derived from the \Ha/\Hb\ ratio. As we describe below (and as it should be expected for IC~10 because of its very low Galactic latitude) the reddening correction is fundamental to get a proper estimation of the line fluxes, and indeed the derived \CHb\ almost always provides values higher than 1.0~dex.

\begin{table*} 
\caption{\label{lines}Dereddened line intensity ratios with respect to $I$(\Hb)=100 for the integrated spectrum of the intense star-forming region [HL90]~111 at the starburst galaxy IC~10 and the regions we specifically analyse within it. Other interesting properties derived from these data are also listed. Colons indicate measurements with an uncertainty larger than 40\%. Double colons indicate a very tentative measurement. Three dots indicate that the emission line has not been detected.}
\begin{tabular}{l l c  c  cc cc c  c}
\hline
$\lambda_0$   &    Ion              &     $f(\lambda)$   & 	  [HL90]~111       &  AP1 (\emph{c})     &  AP2 (\emph{hole})       &    AP3 (\emph{d})      &     AP4 (WR 24)    & Fiber~90  & Fiber~92      \\
\hline

3728.00  &    [\ion{O}{ii}]            &    0.322   &  260$\pm$48       &  171$\pm$25     &   223$\pm$78           &   205$\pm$41      &   158$\pm$60   & 142$\pm$32   & 127$\pm$23   \\
3835.39  & 	H9             &    0.299   &   29.2$\pm$6.7    &   19.0$\pm$4.9  &  	\nodata	      &     \nodata       &   	\nodata	    & \nodata   & \nodata \\ 
3868.75  &  [\ion{Ne}{iii}]            &    0.291   &   36.0$\pm$9.4    &   32.5$\pm$5.4  &  	\nodata	      &    38$\pm$20      &   	\nodata	   & \nodata & 26.6$\pm$9.2     \\
3889.05  & 	H8 +\ion{He}{i}     &    0.287   &   19.2$\pm$5.0    &   20.5$\pm$4.1  &  	\nodata	      &      \nodata      &   	\nodata	      &\nodata  &   \nodata\\
3970.07  & [\ion{Ne}{iii}]+H$\epsilon$  &    0.266   &   28.7$\pm$6.1    &   35.5$\pm$4.3  &  	\nodata	      &    34$\pm$20      &    \nodata    & 32$\pm$19  & 32$\pm$10\\
4101.74  & 	\Hd                    &    0.230   &   43.4$\pm$4.7    &   38.1$\pm$3.5  &  	\nodata	      &    36.3$\pm$8.0   &    \nodata     & \nodata   & 33.3$\pm$8.0 \\
4340.47  & 	\Hg                    &    0.157   &   60.0$\pm$5.0    &   57.5$\pm$4.2  &    62$\pm$26      &    53.7$\pm$7.3   &    59$\pm$11   &   58$\pm$11   & 51.8$\pm$5.9  \\
4363.21  &   [\ion{O}{iii}]            &    0.151   & 	2.42            &    2.9          &     2.0::           &     3.0           &     2.5:          &   2.7:   &   2.9: \\
4471.48  &    \ion{He}{i}              &    0.116   & 	5.0$\pm$1.5     &    5.2$\pm$1.2  &  	\nodata	      &      \nodata      &   	\nodata	    & \nodata &  \nodata  \\
4685.60 &  \ion{He}{ii} neb.       &   0.050    &  \nodata             &  \nodata             &      \nodata             &    \nodata        &       6.3$\pm$2.1 &    14.9$\pm$5.3  & \nodata \\ 
4685.60 &  \ion{He}{ii} WR        &   0.050    &  \nodata             &  \nodata             &      \nodata             &    \nodata        &      89.9$\pm$8.5 &  \nodata &  \nodata \\ 
4861.33  & 	\Hb                    &    0.000   &  100.0$\pm$7.2    &  100.0$\pm$5.8  &   100$\pm$12      &   100.0$\pm$4.6   &   100$\pm$6    &  100$\pm$7  & 100$\pm$4  \\
4958.91  &   [\ion{O}{iii}]            &    -0.026  &  105.2$\pm$5.8    &  125.4$\pm$6.3  &    99$\pm$17      &   123.4$\pm$8.3   &   120$\pm$14    & 126$\pm$12 & 133$\pm$8   \\
5006.84  &   [\ion{O}{iii}]            &    -0.038  &  313$\pm$16       &  373$\pm$18     &   289$\pm$41      &   367$\pm$22      &   344$\pm$30      & 383$\pm$30  & 400$\pm$21 \\
5875.64  &    \ion{He}{i}              &    -0.203 &   14.8$\pm$1.9    &  12.93$\pm$0.87 &    15$\pm$4     &    14.2$\pm$1.7   &    17.1$\pm$4.1    & 12.8$\pm$2.8 & 13.7$\pm$1.1 \\
6548.03  &    [\ion{N}{ii}]            &    -0.296  & 	6.32$\pm$0.72   &   3.91$\pm$0.48 &  	\nodata	      &     5.06$\pm$0.70 &   	\nodata	       & 3.2$\pm$1.3  & 2.90$\pm$0.42\\
6562.82  & 	\Ha                    &   -0.298   &  261$\pm$17       &  273$\pm$20     &   259$\pm$40      &   270$\pm$17      &   267$\pm$22     & 259$\pm$22  & 283$\pm$16 \\
6583.41  &   [\ion{N}{ii}]             &    -0.300  &   20.5$\pm$1.6    &   12.5$\pm$1.1  &    17.3$\pm$5.9   &    16.38$\pm$1.4  &    17.2$\pm$3.1   & 9.8$\pm$2.2 & 9.09$\pm$0.78 \\
6678.15  &    \ion{He}{i}              &    -0.313  & 	3.29$\pm$0.31   &   3.09$\pm$0.29 &  	\nodata	      &     3.00$\pm$0.50 &     4.3$\pm$1.5    & \nodata  & 3.35$\pm$0.31\\
6716.47  &   [\ion{S}{ii}]             &    -0.318  &   16.2$\pm$1.1    &   8.64$\pm$0.70 &    18.7$\pm$4.1   &    13.45$\pm$1.26 &    10.9$\pm$2.5   & 7.1$\pm$1.7  & 6.76$\pm$0.56\\
6730.85  &   [\ion{S}{ii}]             &    -0.320  &   11.94$\pm$0.83  &   6.44$\pm$0.52 &    14.1$\pm$3.4   &     9.5$\pm$1.0   &     6.1$\pm$1.7    & 5.1$\pm$1.3   & 5.10$\pm$0.46\\
7065.28  &    \ion{He}{i}              &    -0.364  & 	2.70$\pm$0.25   &   1.76$\pm$0.23 &  	\nodata	      &     1.34$\pm$0.50 &   	\nodata	    & \nodata    & 1.83$\pm$0.36\\
7135.78  &  [\ion{Ar}{iii}]            &    -0.374  & 	9.85$\pm$0.75   &  10.40$\pm$0.96 &     7.8$\pm$2.8   &    11.0$\pm$1.2   &     9.7$\pm$2.1    & 9.7$\pm$1.7 & 11.08$\pm$0.81\\

\hline

\multicolumn{3}{l}{N. fibres}                         &     254           &     24          &      10           &       11          &         2            &    1     &    1\\
\multicolumn{3}{l}{$F$(\Hb) [10$^{-15}$ erg\,s$^{-1}$\,cm$^{-2}$]}& 219.9$\pm$8.5& 64.6$\pm$2.3 & 1.98$\pm$2.4  &  11.5$\pm$0.53  & 1.123$\pm$0.069 & 1.97$\pm$0.13  & 4.11$\pm$0.17  \\
\noalign{\smallskip}
\multicolumn{3}{l}{\CHb}                               &   1.42$\pm$0.05   & 1.43$\pm$0.08   &   1.43$\pm$0.09   &     1.34$\pm$0.04 &    1.24$\pm$0.04 &  1.15$\pm$0.04  & 1.49$\pm$0.02 \\
\multicolumn{3}{l}{\Wabs\    [\AA]}                  &    2.2$\pm$0.3    &   2.4$\pm$0.3   &   0.6$\pm$0.2     &     1.9$\pm$0.2   &     0.6$\pm$0.1     &  0.9$\pm$0.2 &  2.2$\pm$0.2\\
\multicolumn{3}{l}{$-$\WHa\     [\AA]}                  &   264$\pm$8       &  557$\pm$16     &   69$\pm$6        &  375$\pm$11       &      94$\pm$6   &   97$\pm$7    &  937$\pm$44\\
\multicolumn{3}{l}{$-$\WHb\     [\AA]}                   &    39.8$\pm$2.4   &   88.3$\pm$3.6  &   10.0$\pm$1.3    &  57.7$\pm$3.6     &    11.3$\pm$1.1   &  14.9$\pm$1.3  & 167$\pm$8\\
\multicolumn{3}{l}{$-$\WHg\      [\AA]}                   &    13.81$\pm$0.12 &   28.8$\pm$2.6  &    3.3$\pm$0.9    &    19.3$\pm$1.9   &     3.2$\pm$0.6   & 5.7$\pm$2.3  & 38.4$\pm$1.1\\
\multicolumn{3}{l}{$-$\WHd\       [\AA]}                  &     6.8$\pm$1.1   &  11.4$\pm$1.3   &  \nodata          &     9.9$\pm$1.2   &     \nodata   &   3.02:    &   17.7$\pm$4.5 \\
\hline
\multicolumn{3}{l}{$A_{\rm H\alpha}$ [mag]}         &  2.49$\pm$0.09  &   2.51$\pm$0.14  &  2.51$\pm$0.16  &   2.35$\pm$0.07 &   2.18$\pm$0.07 & 2.02$\pm$0.07 & 2.61$\pm$0.04 \\
\multicolumn{3}{l}{$I$(\Ha) [10$^{-14}$ erg\,s$^{-1}$\,cm$^{-2}$]}&  1547$\pm$45&   485$\pm$19     &   14.5$\pm$0.9   &  70.0$\pm$2.4   &   22.3$\pm$1.1    & 7.55$\pm$0.42 & 36.1$\pm$2.4\\
\multicolumn{3}{l}{$L$(\Ha) [10$^{36}$ erg\,s$^{-1}$]}   & 1185$\pm$64  &  371$\pm$34      &   11.1$\pm$1.3   &    53.6$\pm$4.5  & 17.1$\pm$1.5 & 5.78$\pm$0.54 & 27.6$\pm$2.4 \\
\multicolumn{3}{l}{\MHII\ [\Mo]}                                  & 17600$\pm$950   &  5510$\pm$510  &   165$\pm$19     &    796$\pm$67    &  254$\pm$22  &  86$\pm$8 &  410$\pm$36\\
\multicolumn{3}{l}{\MHII/area\ [\Mo\,pc$^{-2}$]}       & 4.60$\pm$0.25   &  15.3$\pm$1.4   &   1.10$\pm$0.13  &   4.81$\pm$0.40 &  8.43$\pm$0.73  &  5.71$\pm$0.53 &  27.2$\pm$2.4\\



\hline
\end{tabular}
\end{table*}

Unfortunately, the \ion{Hg}{i}~$\lambda$4358 telluric emission was completely blended with the faint [\ion{O}{iii}] $\lambda$4363 auroral line, which has a redshifted wavelength of $\lambda$4359 in IC~10. 
Using the regions with the faintest ionized gas emission (AP~2, see below), we have estimated an average value of the \ion{Hg}{i}~$\lambda$4358 line flux of 1.8$\times 10^{-16}$~erg~cm$^{-2}$~s$^{-1}$ for each spaxel. In these resolution elements, and assuming an electron temperature of $T_e\sim$10000~K, we estimate that the  [\ion{O}{iii}]~$\lambda$4363 line flux is $\sim$80--100 times fainter than that measured for the \ion{Hg}{i}~$\lambda$4358 line flux and hence completely negligible. However, the contribution of the  [\ion{O}{iii}] $\lambda$4363 line to the \ion{Hg}{i}~$\lambda$4358 line flux is important in the brightest regions of the nebula. In these cases, and applying the same method, we estimate that $F$([\ion{O}{iii}]~$\lambda$4363)$\sim$0.4--0.6$\times F$(\ion{Hg}{i}~$\lambda$4358). The contribution of the auroral line is even more important when integrating the line fluxes of several fibers, as we do in several defined apertures. Hence, we will use the derived average value of the \ion{Hg}{i}~$\lambda$4358 line flux to get a estimate of the flux of the [\ion{O}{iii}]~$\lambda$4363 line within these areas.

  \begin{figure*}
          \begin{tabular}{cc}
   \includegraphics[width=0.4\linewidth,angle=90]{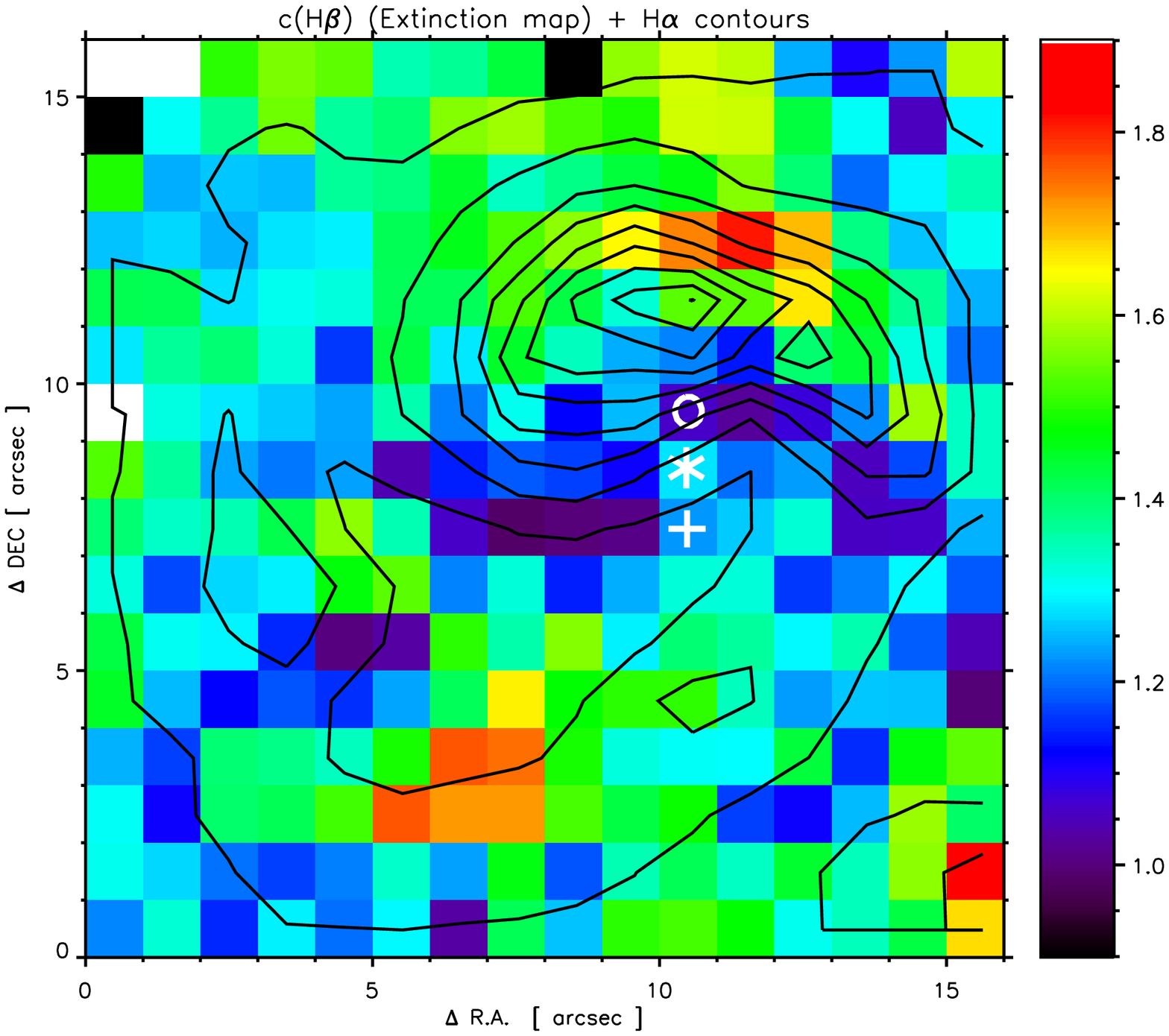}  &
     \includegraphics[width=0.4\linewidth,angle=90]{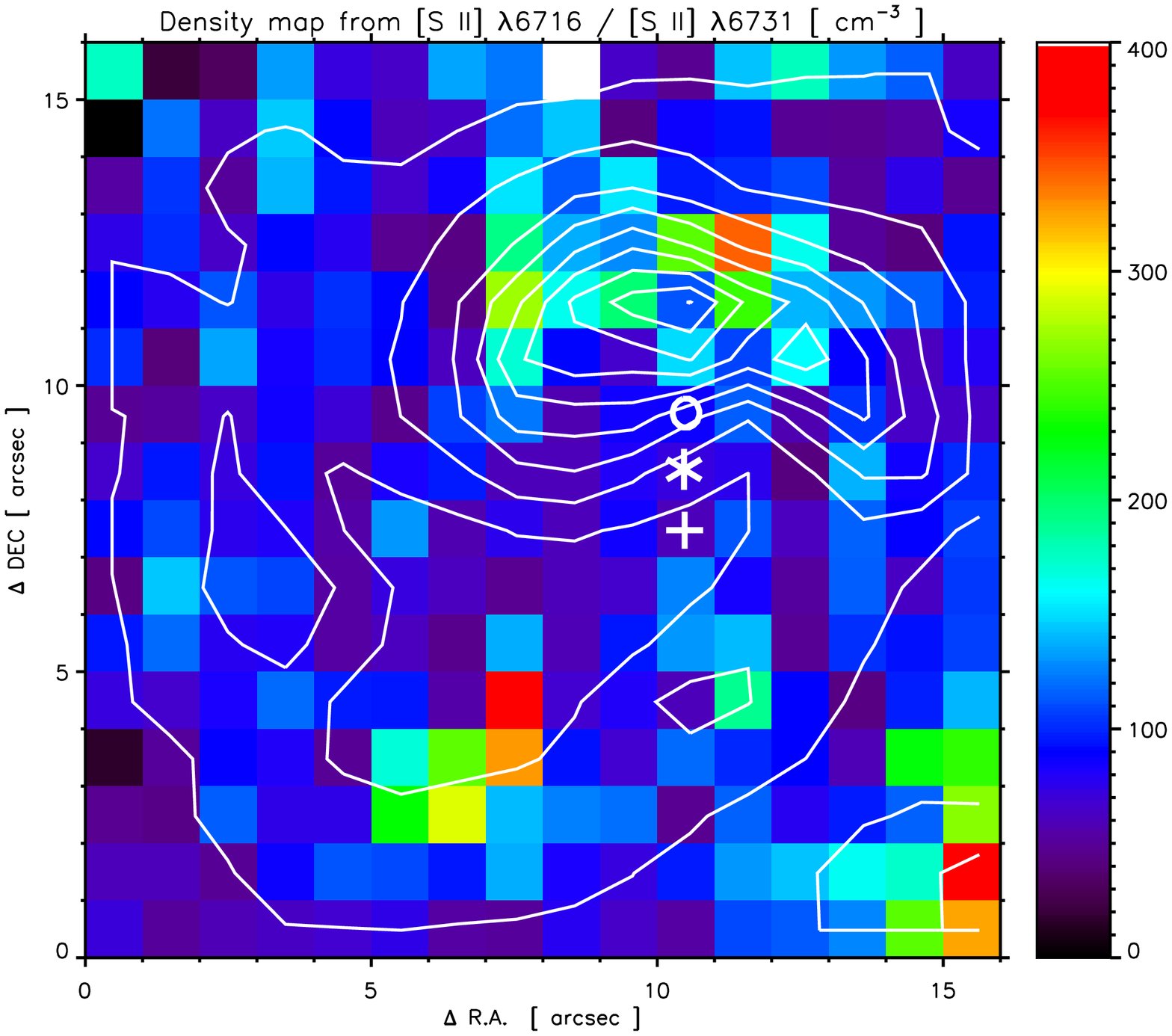} \\
            \end{tabular}
   \caption{\label{chb} (\emph{Left}) Map of the reddenning coefficient, \CHb, derived from the \Ha/\Hb\ ratio in the field observed in IC~10.  (\emph{Right}) 
 Density map derived from the  [\ion{S}{ii}] $\lambda$6716/$\lambda$6731 ratio. 
 The H$\alpha$ contours are overlaid in both maps. Symbols are the same that in Fig.~\ref{halpha}.}
  \end{figure*}

\section{Results} \label{resul}

The reconstruction of the \Ha\ map using our 2D PMAS spectroscopy is shown in Fig.~\ref{halpha}. This map corresponds to the area named as [HL90]~111 by \cite{HL90}. We distinguish a very strong star-forming region at the NW (region \emph{c}), which has a size of $\sim10\arcsec\times6\arcsec$. Other fainter \HII\ regions are located at the E (knot \emph{d}) and in the SW corner (knot~\emph{a}), this last one is only partially covered by our FoV (see Fig.~\ref{mapa}). An important feature is a hole located at the south of the brightest star-forming region. Indeed, there is an arc-like filament of ionized gas \citep[knot~\emph{e} following][]{HL90} connecting the SW tip of the brightest star-forming area (\emph{c}) with the \HII\ region located at the east (\emph{d}). The minimum of \Ha\ emission (the \emph{hole}) is located between these three structures.

We have integrated the spectra within several regions to analyse in detail the properties of the ionized gas and to maximize the signal-to-noise (S/N) ratio of the 1D-spectra. Figure~\ref{apertures} shows the position of these regions: AP~1 (which corresponds to the brightest star-forming region, [HL90]~111c), AP~2 (which corresponds to the \emph{hole} of \Ha\ emission), AP~3 (which corresponds to [HL90]~111d) and AP~4 (which integrates the spectra of the two spaxels where the WR features are detected). For these four regions we measured all the line fluxes following the same procedure explained before. We also carefully analysed two individual fibers: \#90 (spaxel [11,10]), where a tentative detection of the nebular \ion{He}{ii}~$\lambda$4686 line is found, and \#92 (spaxel [11,13]), which corresponds to the maximum in the \Ha\ emission and hence to the fiber where emission lines have the best S/N ratio.

Because of the stacking of the spectra, some emission lines that were not easily observed in the individual spa\-xels (i.e., H9, H8, \Hd, [\ion{Ne}{iii}] $\lambda$3869,  [\ion{Ne}{iii}] $\lambda$3967+H$\epsilon$, \ion{He}{i}~$\lambda$4471) are now clearly distinguished. That allowed us to perform a more accurate determination of the reddening coefficient within these areas. We considered all available \HI\ Balmer line ratios (\Ha/\Hb, \Hg/\Hb, \Hd/\Hb, and also H8/\Hb\ in the brightest region) and applied the formalism explained in \cite{LS06} and \cite{LSE09} to simultaneously derive \CHb\ and the equivalent width of the absorption in the \HI\ lines, \Wabs, which we assumed to be same for all \HI\ Balmer lines. Table~\ref{lines} shows the dereddened line intensity ratios and their associated errors, as well as the adopted $f(\lambda)$ of each emission line, measured for the total integrated spectrum, the four regions we have carefully analyzed, and fibers \#90 and \#92. In this table we also include other important quantities such as the size of the extracted aperture,  the observed \Hb\ flux (uncorrected for extinction), the adopted values of \CHb\ and \Wabs, and the equivalent widths of the \HI\ Balmer lines. Colons indicate errors of the order or larger than 40\%. As explained before, the contribution of the \ion{Hg}{i}~$\lambda$4358 line flux to the [\ion{O}{iii}] $\lambda$4363 line flux was subtracted assuming the average value of the \ion{Hg}{i}~$\lambda$4358 line and the number of fibers used to get the integrated spectrum of the region.

   \begin{figure*}
   \centering
     \begin{tabular}{cc}
   \includegraphics[width=0.42\linewidth,angle=90]{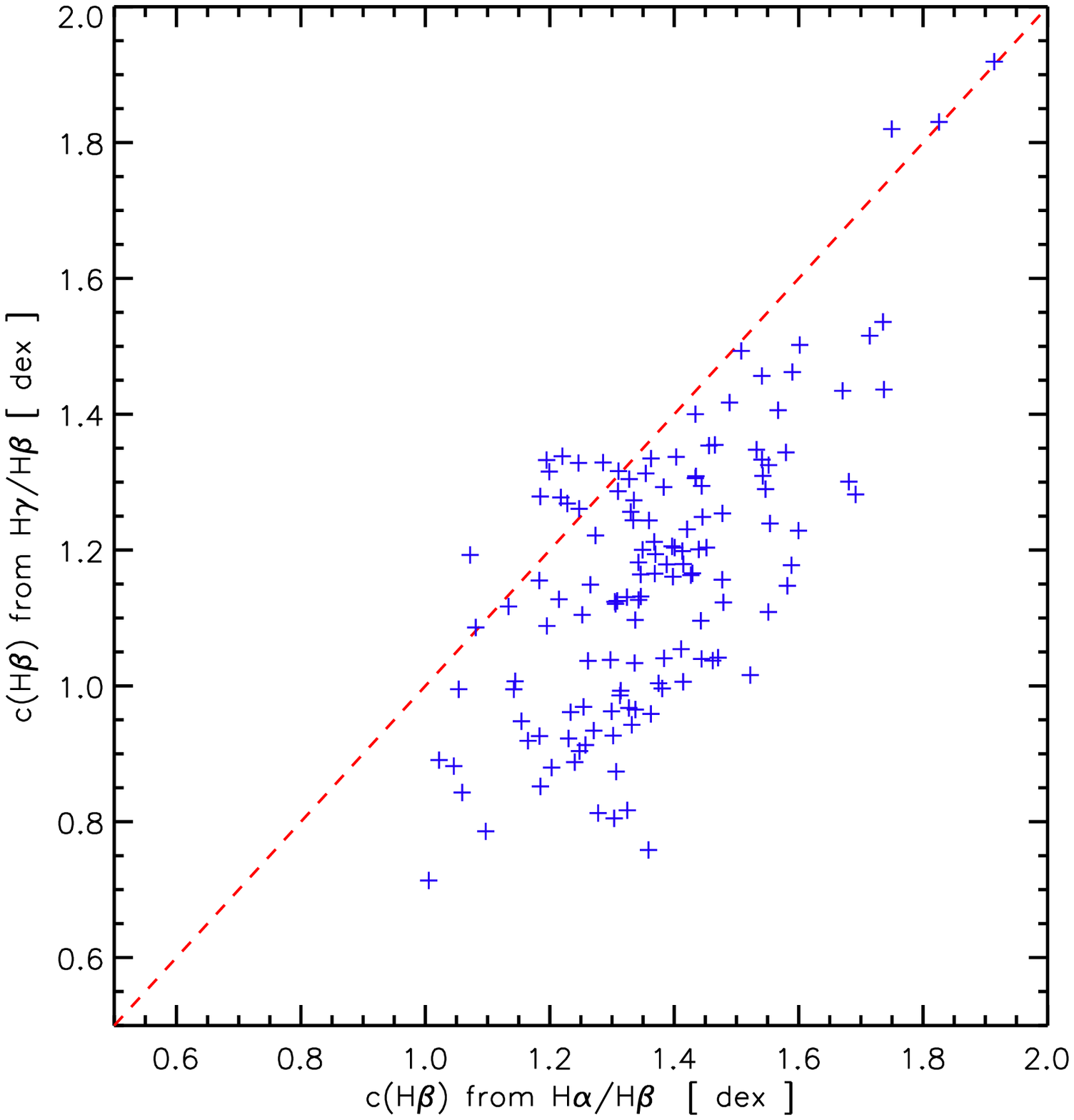} & 
      \includegraphics[width=0.42\linewidth,angle=90]{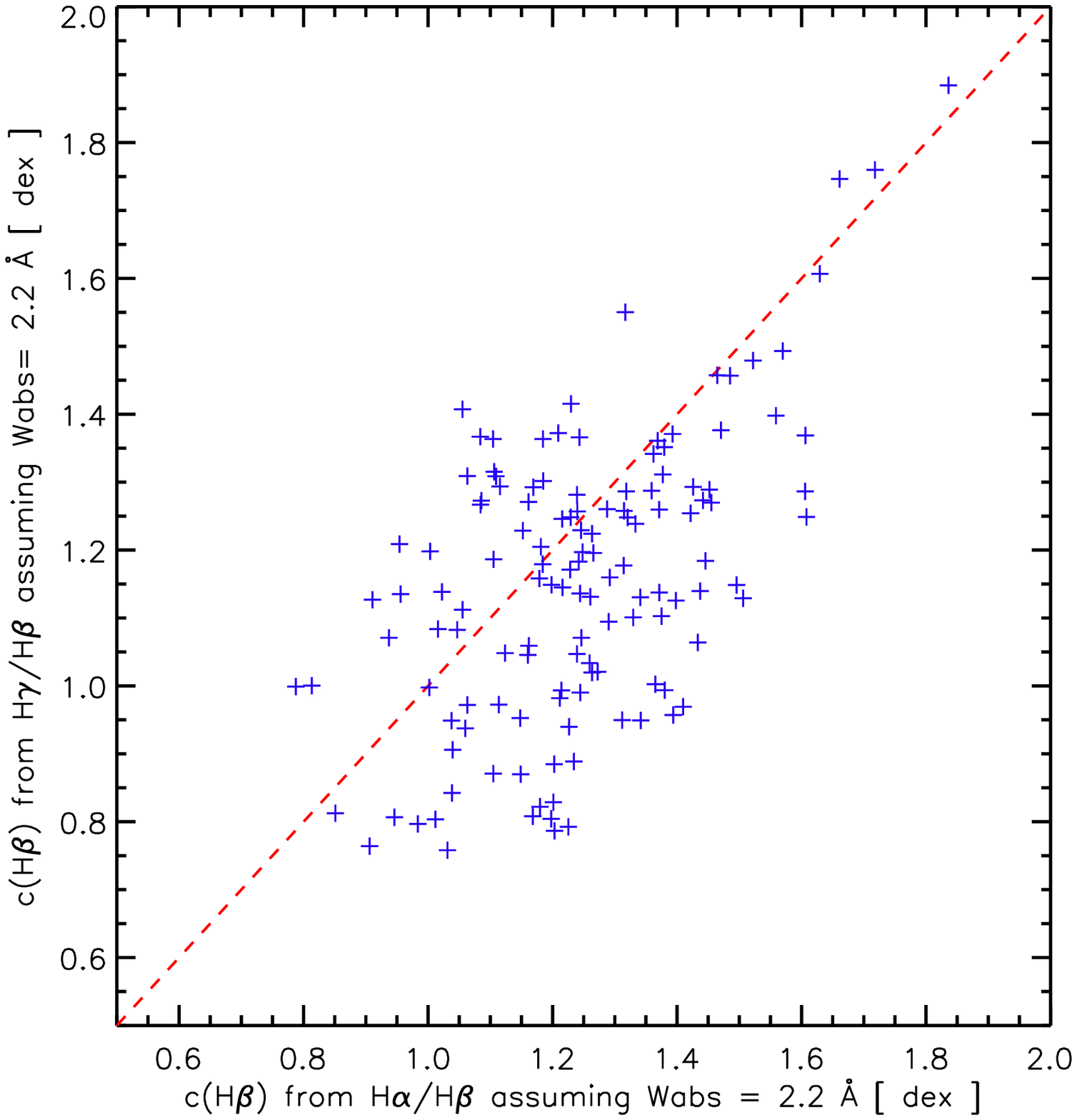} \\ 
            \end{tabular}
   \caption{   \label{hahbhg} Comparison between the \CHb\ derived from the \Ha/\Hb\ ratio (x-axis) and that derived from the H$\gamma$/\Hb\ ratio (y-axis). The $x=y$ is plotted with a discontinuous red line. (\emph{Left diagram}) \CHb\ derived no considering \HI\ Balmer line absorption. (\emph{Right diagram}) \CHb\ derived assuming an absorption of \Wabs=2.2~\AA\ in the \HI\ Balmer lines.} 
 \end{figure*}

The \Ha\ map obtained using 2D spectroscopy allows us to determine the total luminosity and the mass of the ionized gas within these regions. The extinction-corrected \Ha\ flux was determined considering 
\begin{equation}
I({\rm H\alpha}) = F({\rm H\alpha})  \times10^{c({\rm H}\beta)\times [ f({\rm H}\alpha) +1]}  =  F({\rm H\alpha})\times10^{A_{\rm H\alpha}/2.5}.
 \end{equation}
The value of the extinction in the \Ha\ line, $A_{\rm H\alpha}$, was computed in each region considering the value of the $f$(\Ha) provided by the \cite{Cardelli89} extinction law and the \CHb\ derived in our analysis, $A_{\rm H\alpha}$ = 1.755$\times$\CHb, and it is tabulated in Table~\ref{lines}. The total \Ha\ luminosity was computed assuming a distance to IC~10 of 0.8~Mpc, the uncertainty of $L$(\Ha) also includes the uncertainty in the distance. We estimate a total \Ha\ luminosity of (1.185$\pm$0.064)$\times$~$10^{36}$~erg\,s$^{-1}$. Using the \cite{Calzetti07} calibration between $L$(\Ha) and the star-formation rate (SFR) we derive \mbox{$SFR_{\rm H\alpha}=(6.28\pm0.34)\times10^{-3}$~\Moy.} This value corres\-ponds to a SFR density of \mbox{$1.64\times0.09$~\Moy\,kpc$^{-2}$,} and confirms the starburst activity that the center of IC~10 is experiencing.

Finally, the mass of the ionized gas, \MHII, was computed using the expression provided by \cite{enrique02},  \MHII~=~1.485$\times 10^{-35} L$(\Ha)$\times$(\Ne/100), where \Ne\ is the electron density (in cm$^{-3}$) derived from our spectroscopic data (which was always 100~\cmc) and the result is obtained in solar masses. The derived values of $L$(\Ha) and \MHII\ of each region are also included in Table~\ref{lines}. We estimate a total ionized gas mass of $(1.76\pm0.10)\times10^{5}$~\Mo\ within all the observed field. Table~\ref{lines} also lists the ionized gas mass per surface area, \MHII/area. 
It has a minimum value of  1.10~\Mo\,pc$^{-2}$ in the \emph{hole} but it reaches the maximum values in the strong star-forming region, which has \MHII/area=15.3~\Mo\,pc$^{-2}$.

  \begin{figure*} 
   \centering
    \begin{tabular}{cc}
 \includegraphics[width=0.4\linewidth,angle=90]{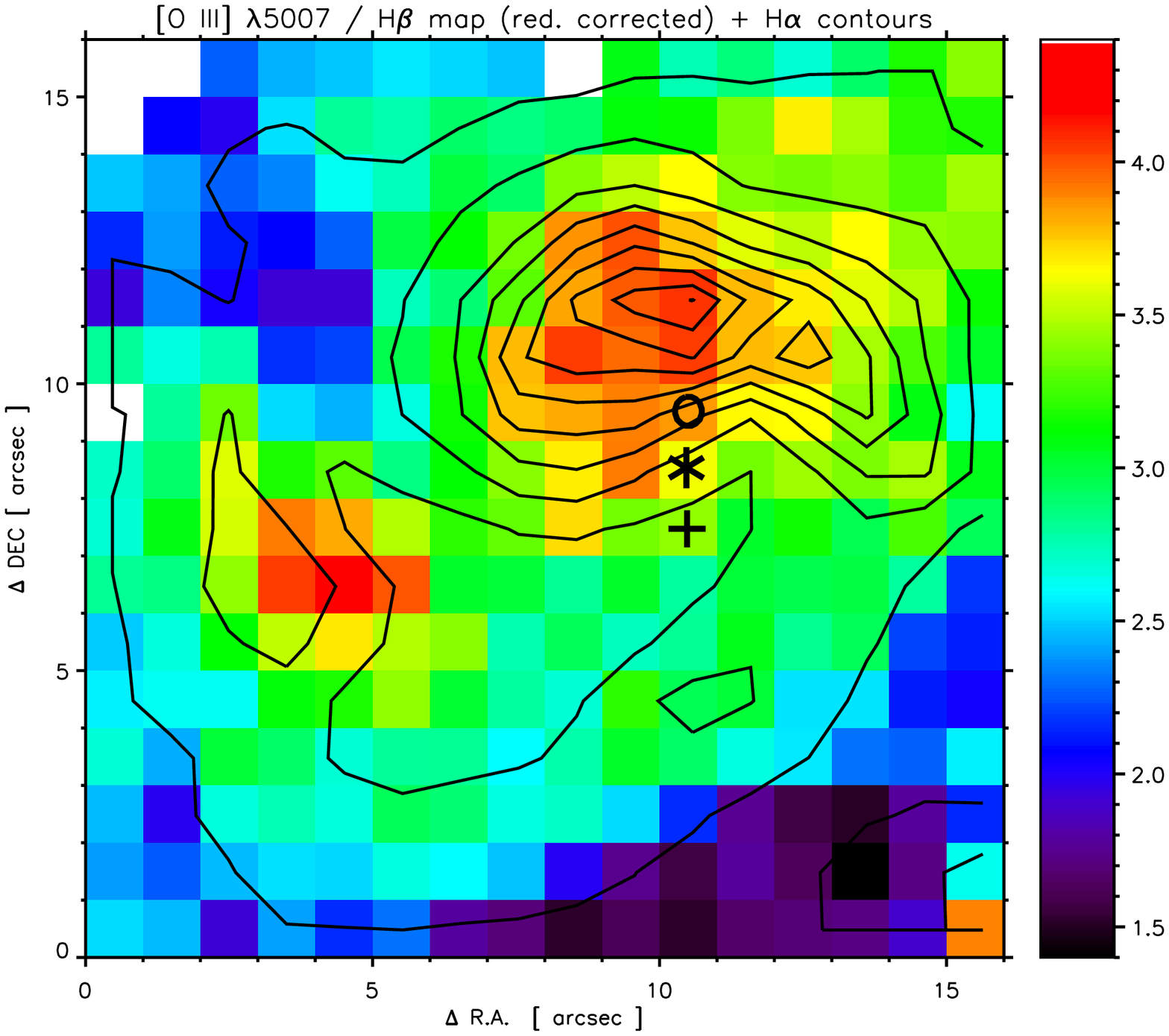} &
 \includegraphics[width=0.4\linewidth,angle=90]{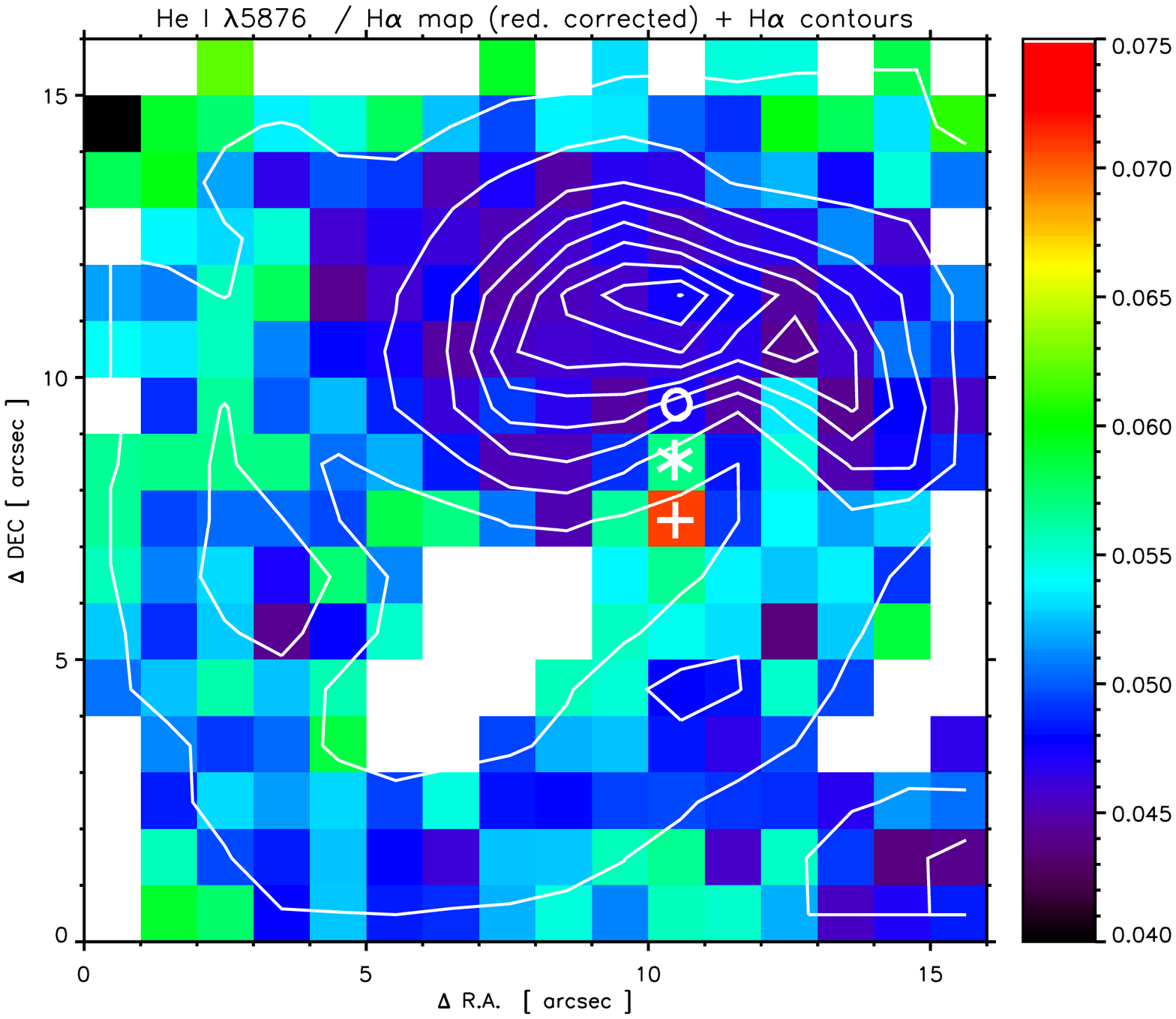} \\ 
 \includegraphics[width=0.4\linewidth,angle=90]{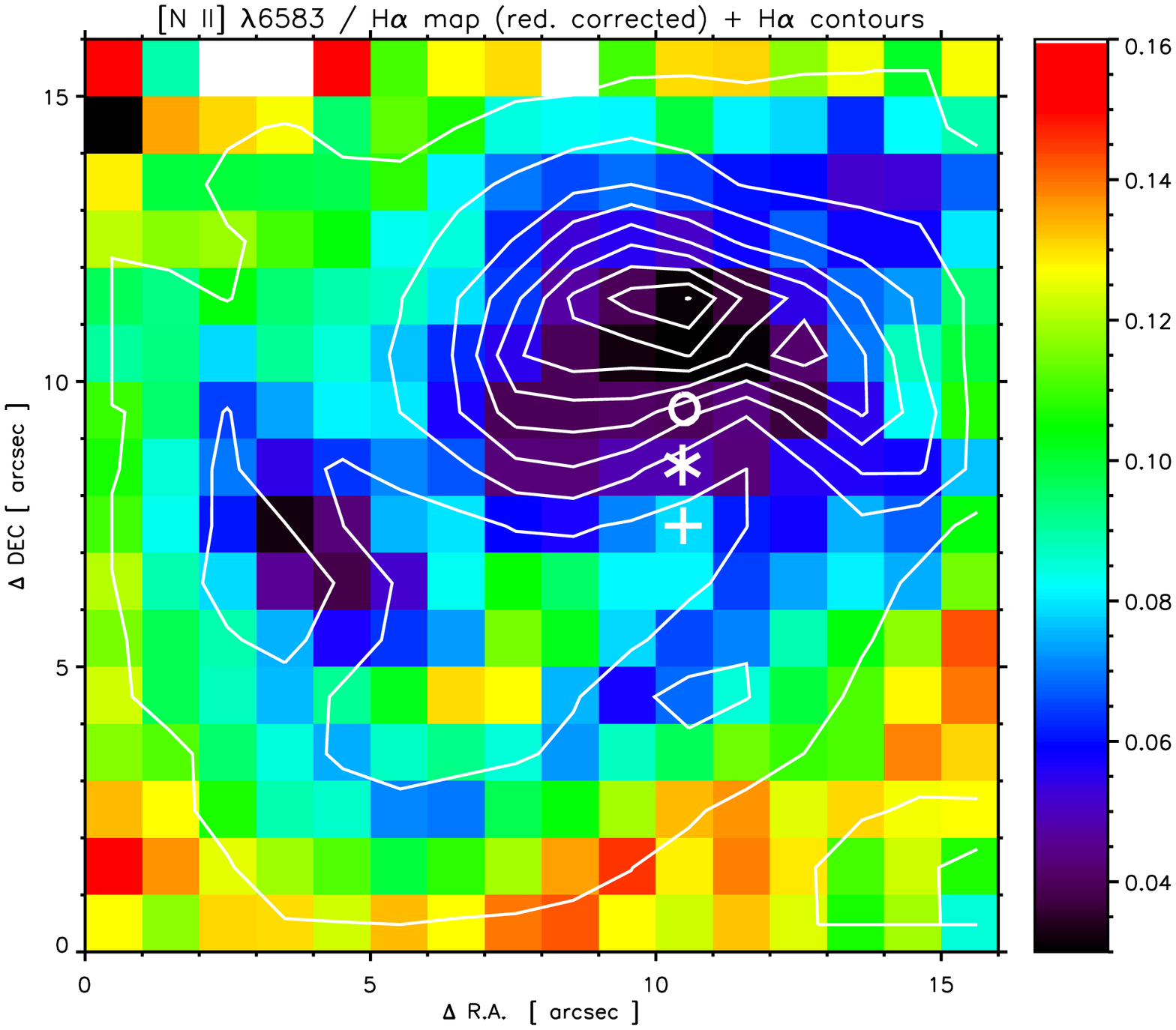} &
 \includegraphics[width=0.4\linewidth,angle=90]{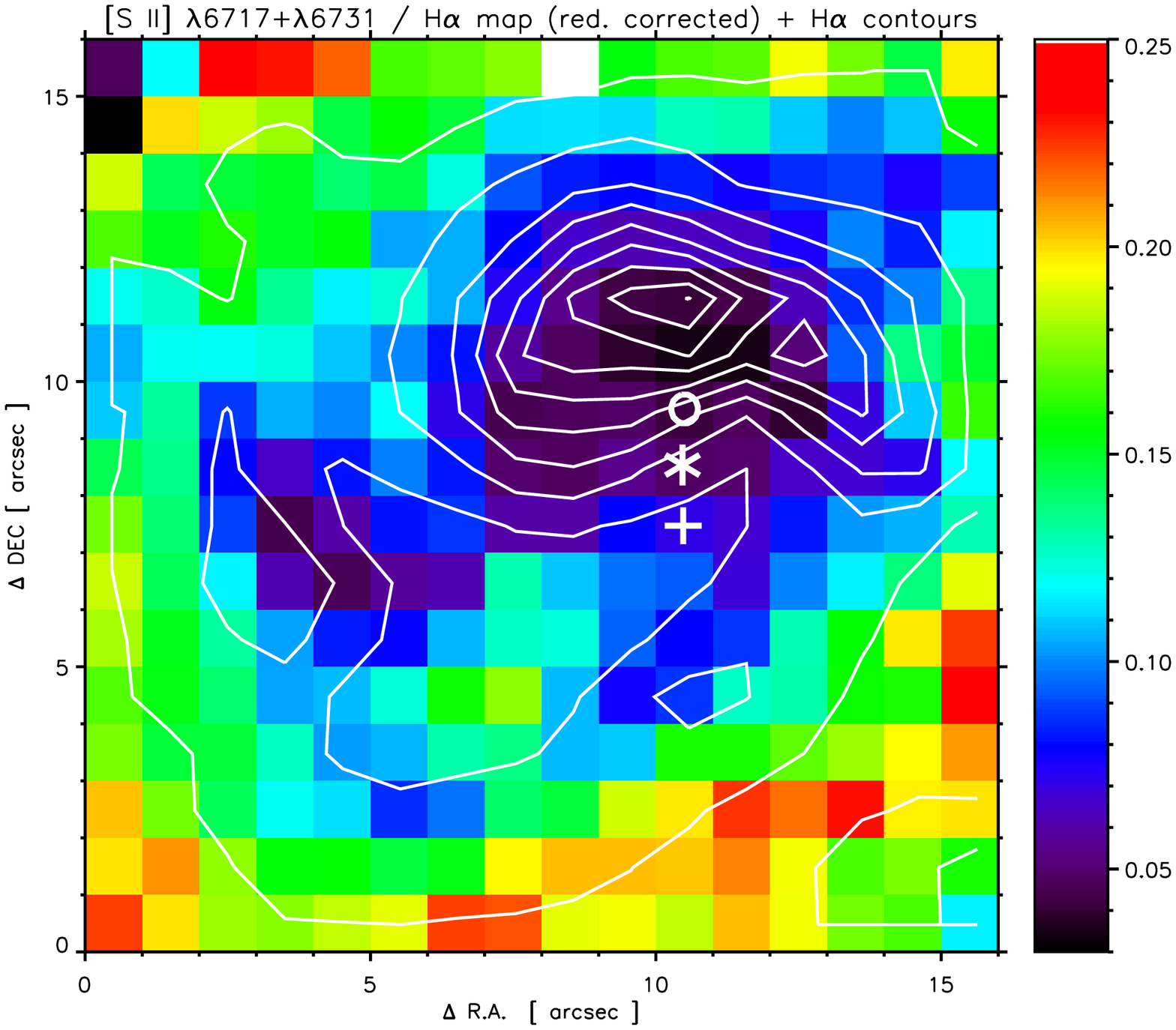} \\

 \end{tabular}
   \caption{   \label{mapas} Emission line ratio maps with H$\alpha$ contours overlaid into the IC~10 PMAS field
            of: (\emph{tot left}) [\ion{O}{iii}]~$\lambda$5007/\Hb,
            (\emph{top right})   \ion{He}{i}~$\lambda$5876/\Ha,
               (\emph{bottom left}) [\ion{N}{ii}]~$\lambda$6583/\Ha,
               (\emph{bottom right})   [\ion{S}{ii}]~$\lambda\lambda$6716,6731/\Ha.
                All maps have been corrected for reddening using the \CHb\ value derived from the \Ha/\Hb\ ratio.  Symbols are the same that in Fig.~\ref{halpha}.}
 \end{figure*}

 \subsection{Reddening map}
 
Left panel of Fig.~\ref{chb} shows the map of the reddening coefficient derived in IC~10 using the \Ha/\Hb\ ratio. \CHb\ has values between 1.0 and 1.8~dex. These values correspond to an extinction in the \mbox{$V$-band,} $A_V$, between 2.15 and 3.86~mag or, equivalently, a color excess, $E(B-V)$, between 0.69 and 1.25~mag. The integrated value of the  extinction in the \mbox{$V$-band} is $A_V=3.05\pm0.11$~mag. These values nicely match previous values of the extinction given in the literature (e.g., Lee, Freedman \& Madore 1993; Vacca et al. 2007; Lozinskaya et al. 2009).
These values confirm the high extinction existing in the direction to IC~10, but because of its very low Galactic latitude we know that a very important fraction of the observed extinction is due to the Galactic ISM. In any case, assuming that the Galactic extinction is mainly constant in all the observed region, we appreciate that the dust within the star-forming region has a rather inhomogeneous distribution. In particular, we note the the lower values of \CHb\ are located near the WR stars, at the SE of the brightest \HII\ region, that show \CHb\ between 1.0 and 1.3~dex. The maximum values of the reddening coefficient are found at the NW of the brightest star-forming region, which has values between 1.5 and 1.8~dex. Other areas (at the south of the \emph{hole} and in the knot~\emph{a} at the SW corner) also show a high extinction. The average value we determine for all the observed region is \CHb=1.42$\pm$0.05~dex. This peculiar distribution suggests that the dust around the massive stars has been evacuated by interstellar winds and/or radiation pressure or partially destroyed. Although this effect has not been investigated in detail, there are indications of decreasing values of the extinction around young stars in the Orion Nebula Cluster \citep{Hillenbrand97}.
 
We compared the reddening coefficient derived from the \Ha/\Hb\ ratio with that estimated using the \Hg/\Hb\ ratio in those fibers in which \Hg\ is properly measured (52\% of the spaxels). The left panel of Fig.~\ref{hahbhg} plots the \CHb\ values derived in each case. As we see, although the correspondence between both values is relatively good, the \CHb\ derived from the \Hg/\Hb\ ratio is systematically lower than that computed using the \Ha/\Hb\ ratio. This effect is consequence of the \HI\ Balmer line absorption of the underlying stellar population. The right panel of Fig.~\ref{hahbhg} plots the \CHb\ values derived in each case when considering a \HI\ Balmer line absorption of \Wabs=2.2~\AA, the value we derived iteratively for the integrated spectrum of the observed field (see Table~\ref{lines}). Indeed, the correspondence between both \CHb s is much better when \Wabs\ is taken into account, although the dispersion of the data is slightly higher because of the higher uncertainty in both the \Hg\ fluxes and equivalent widths.

\subsection{Physical conditions of the ionized gas} \label{condpmas}    

Because of the faintness of the auroral lines and the conta\-mination of the \ion{Hg}{i}~$\lambda$4358 telluric line, it was not possible to derive a temperature map from our PMAS data. We used the [\ion{S}{ii}]~$\lambda$6716/$\lambda$6731 ratio to derive a bidimensional map of \Ne, assuming a \Te=10500\,K (the electron temperature derived for the ionized gas, see below).
Lower values of the [\ion{S}{ii}] $\lambda$6716/$\lambda$6731 ratio indicate higher values in \Ne. We used the {\sc temden} task of the  {\sc nebular} package of {\sc iraf} \citep{SD95} with updated atomic data \citep[see][]{garciarojasetal05}. 
This map is shown in the right panel of Fig.~\ref{chb}.
Purple and blue points indicate 
the places where the electron density is in the low limit,  \Ne$\lesssim$100\,\cmc\ ([\ion{S}{ii}] $\lambda$6716/$\lambda$6731 ratios between 1.40 and 1.32). 
As we see, these colors are the predominant in the observed region. However, we note some places with 
electron densities between 100 and 400\,\cmc ([\ion{S}{ii}] $\lambda$6716/$\lambda$6731 ratios between 1.32 and 1.10). Hence, these regions have a higher gas density than that found in the surrounding ISM. It is important to note that the three regions with the highest electron density are precisely located in the places where we find the maximum of the reddening coefficient (left panel in Fig.~\ref{chb}).

For the integrated spectra, we derived the flux of the auroral [\ion{O}{iii}]~$\lambda$4363 line by subtracting the expected flux of the telluric \ion{Hg}{i}~$\lambda$4358 line considering the number of fibers used in each region. We then applied the {\sc temden} task of the  {\sc nebular} package of {\sc iraf} to derive the \Te[\ion{O}{iii}] in each region. For AP~2, which integrates the flux of the faintest regions of the nebula, the value derived for the [\ion{O}{iii}]~$\lambda$4363 line is just a residual value. In all cases, \Te[\ion{O}{iii}] is around 10500~K, in agreement with previous results found in IC~10 \cite[i.e.,][]{Garnett90, Crowther03, Lozinskaya09}, so we are quite confident in the validity of our decontamination of \ion{Hg}{i} emission. Using the relation between  \Te[\ion{O}{iii}]  and  \Te[\ion{O}{ii}] provided by \cite{Garnett92},   \mbox{\Te[\ion{O}{ii}]=0.7$\times$\Te[\ion{O}{iii}]+3000,} we have estimated the electron temperature for the low-ionization potential ions. We compile the electron temperatures derived in each of the regions in Table~\ref{abundancias}. This table also lists the electron density computed for each region using the  [\ion{S}{ii}]~$\lambda$6716/$\lambda$6731 ratio. In all cases, this ratio provided a density which was  below the low-density limit, \Ne$<$100~\cmc, and hence we always considered \Ne=100~\cmc.

 \subsection{Nature of the ionization} \label{excitation}       
  
Figure~\ref{mapas} shows the emission line ratio maps we obtained for the analyzed region in IC~10. All maps have been corrected for reddening using the \CHb\ value derived from the \Ha/\Hb\ ratio. 
We show the maps corresponding to  [\ion{O}{iii}]~$\lambda$5007/\Hb, 
[\ion{N}{ii}]~$\lambda$6583/\Ha, ([\ion{S}{ii}]~$\lambda$6716+$\lambda$6731)/\Ha,
and \ion{He}{i}~$\lambda$5876/\Ha. White colors indicate regions in which the studied emission line was not detected or it was too noisy.
  
\begin{figure}
\centering
 \includegraphics[width=0.85\linewidth,angle=90]{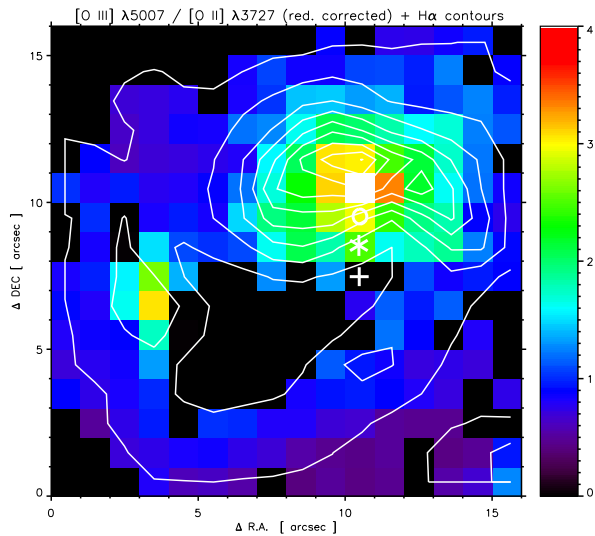}
  \caption{   \label{o3o2} Map of the dereddened [\ion{O}{iii}]~$\lambda$5007~/~[\ion{O}{ii}]~$\lambda$3727 ratio computed in the intense star-forming region  [HL90]~111 within the starburst galaxy IC~10. The contours of the \Ha\ emission are overlaid.  Symbols are the same that in Fig.~\ref{halpha}.}
 \end{figure}

As we see, the maps are showing the ionization structure of the giant \HII\ regions in the center of IC~10. That is specially evident in the brightest star-forming region [HL90]~111c and in the \HII\ region located at the SE, [HL90]~111d, both show the maxima of the [\ion{O}{iii}]~$\lambda$5007/\Hb\ ratio at their centers, while the maxima of the low-ionization potential ion ratios is located surrounding those central bright areas. This effect is clearly noticed in the  [\ion{O}{iii}]~$\lambda$5007~/~[\ion{O}{ii}]~$\lambda$3727 ratio map plotted in Fig~\ref{o3o2}, which shows the higher values at the center of the strong star-forming regions. This result is very important and should be taken into account, as we will see below, when using empirical calibrations to get the chemical abundance of the ionized gas. It is also interesting to note that the highest values of the [\ion{O}{iii}] $\lambda$5007/\Hb\ ratio is close to the maximum of the \Ha\ emission, but they are not completely coincident.

We have also analyzed the diagnostic diagrams plotting two different excitation line ratios for classifying the excitation mechanism of the ionized gas, as initially proposed by \cite{VO87}. Figure~\ref{diagnostico} plots the typical [\ion{O}{iii}]~$\lambda$5007/\Hb\ versus  [\ion{N}{ii}]~$\lambda$6583/\Ha\ and  [\ion{O}{iii}]~$\lambda$5007/\Hb\ versus  ([\ion{S}{ii}]~$\lambda$6716+$\lambda$6731)/\Ha\ diagrams for all the spaxels for which we have a good measurement of these lines. In these diagnostic diagrams, \HII\ regions and starburst galaxies lie within a narrow band, but when the gas is excited by shocks, accretion disks, or cooling flows (in the case of AGNs or LINERs) its position is away from the locus of \HII\ regions. We used the analytic relations given by the photoionized models provided
by \cite{Dopita00} for extragalactic \HII\ regions (that assume instantaneous star-formation within star-forming regions) and the \cite{Kewley01} models for starburst galaxies (which consider conti\-nuous star formation and more realistic assumptions about the physics of starburst galaxies). In both cases, photoionization models with very different metallicities (between 0.1 and 3~$Z_{\odot}$) and ionization parameters lie always under these dividing lines. Indeed, the dividing line given by the \cite{Kewley01} models represents an upper envelope of positions of star-forming galaxies. 
The left panel of Fig.~{\ref{diagnostico} includes the empirical relation between the  [\ion{O}{iii}]~$\lambda$5007/\Hb\ and the [\ion{N}{ii}]~$\lambda$6583/\Ha\ provided by \cite{Kauffmann03} analysing a large data sample of star-forming galaxies from the Sloan Digital Sky Survey \citep[SDSS;][]{York00}.

\citet{Kewley+06} suggested that those objects between the theoretical line computed by \cite{Kewley01} and the empirical line found by \cite{Kauffmann03} may be ionized by both massive stars and shocks.
However, \cite{PerezMonteroContini09} showed that objects located in this area may also be pure star-forming galaxies but with a high N content. In our case, all data points (even those spaxels for which a possible N enrichment has been detected, see \S5.1) lie within this area, but below the \cite{Kewley01} theoretical line. This clearly indicates 
that photoionization is the main excitation mechanism of the gas and that there is very little evidence for a significant contribution from shock excitation. The position of the data points in the  [\ion{O}{iii}]~$\lambda$5007/\Hb\ versus  [\ion{N}{ii}]~$\lambda$6583/\Ha\ diagram perfectly matches with the predictions given by the \cite{Dopita00} models. In the case of the [\ion{O}{iii}]~$\lambda$5007/\Hb\ versus   ([\ion{S}{ii}]~$\lambda$6716+$\lambda$6731)/\Ha\ diagram, the data points also agree well with these models, but the data tend to move to the predictions of the upper envelope given by  \cite{Kewley01} in points with log([\ion{O}{iii}]~$\lambda$5007/\Hb)$<$0.5 and log([\ion{S}{ii}]~$\lambda$6716+$\lambda$6731)/\Ha$>-0.8$.

     \begin{figure}
   \centering
   \includegraphics[width=0.85\linewidth,angle=90]{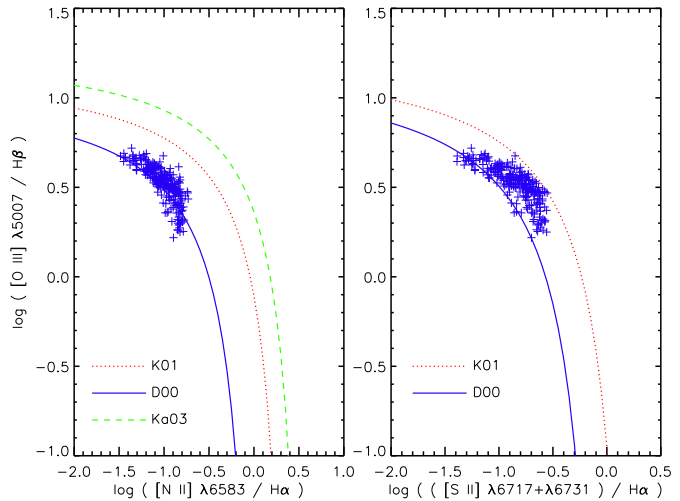} 
   \caption{\label{diagnostico}Comparison of some observational flux ratios obtained for all available spaxels with the diagnostic diagrams proposed by Dopita et al. (2000), blue continuous line (D00), and
Kewley et al. (2001), red discontinuous line (K01). The left panel also shows the empirical relation provided by Kauffmann et al. (2003) with a dotted-dashed dark yellow line (Ka03).} 
  \end{figure}

\begin{table*} 
\caption{\label{abundancias}Physical conditions and chemical abundances of the ionized gas for the regions analyzed in IC~10 using our PMAS data.  Three dots indicate that the abundance could not be derived because of the non-detection of some emission lines. }
\begin{tabular}{l   c  cc cc  c   c}
\hline
   Region                           & 	  [HL90] 111       &  AP1 (\emph{c})     &  AP2 (\emph{hole})       &    AP3 (\emph{d})      &     AP4 (WR 24)     &     Fiber~90          &    Fiber~92\\
\hline
       
\Te\ [\ion{O}{iii}] [K]          &   10600$\pm$600   &    10600$\pm$600 &  10500$\pm$1000 &  10700$\pm$800  &   10500$\pm$1000 &  10500$\pm$1000 &  10500$\pm$800 \\
\Te\ [\ion{O}{ii}]  [K]          &   10400$\pm$400   &    10400$\pm$400 &  10400$\pm$700   &  10500$\pm$600  &   10400$\pm$700   &   10400$\pm$700 &   10400$\pm$600 \\
\Ne\ [cm$^{-3}$]               &      100                    &   	 100	     	         &     100	                   &	  100                	  &      100	                    &      100	           &   100 \\
\noalign{\smallskip}
12+log(O$^+$/H$^+$)     &     7.96$\pm$0.12	 &   7.78$\pm$0.10 	&  7.89$\pm$0.24    &  7.84$\pm$0.12    &    7.74$\pm$0.19       &    7.70$\pm$0.20  &   7.65$\pm$0.13 \\
12+log(O$^{++}$/H$^+$) &     7.96$\pm$0.06	 &   8.03$\pm$0.07 	&  7.94$\pm$0.10    &  8.01$\pm$0.07    &    8.02$\pm$0.08       &    8.05$\pm$0.09  &   8.07$\pm$0.08   \\
12+log(O/H)                     &     8.26$\pm$0.09	 &   8.22$\pm$0.08 	&  8.22$\pm$0.17    &  8.23$\pm$0.09    &    8.20$\pm$0.12       &    8.21$\pm$0.13  &   8.21$\pm$0.09    \\
\noalign{\smallskip}
 log (O$^{++}$/O$^+$)    &     0.00$\pm$0.13 &   0.25$\pm$0.12 	&  0.04$\pm$0.25    &  0.17$\pm$0.14    &    0.28$\pm$0.20       &    0.36$\pm$0.21  &   0.42$\pm$0.15   \\
 12+log(N$^+$/H$^+$)    &     6.54$\pm$0.05 &   6.33$\pm$0.05 	&  6.49$\pm$0.15    &  6.43$\pm$0.05    &    6.49$\pm$0.10       &    6.24$\pm$0.11  &   6.19$\pm$0.06 \\
12+log(N/H)                     &     6.84$\pm$0.08	 &   6.78$\pm$0.09 	&  6.81$\pm$0.20    &  6.90$\pm$0.10    &    6.95$\pm$0.17       &    6.75$\pm$0.18  &   6.76$\pm$0.12   \\
log(N/O)                          &    -1.42$\pm$0.12	 &  -1.45$\pm$0.11 	& -1.41$\pm$0.24    & -1.41$\pm$0.13    & {\bf-1.26$\pm$0.20} &  -1.46$\pm$0.21  &  -1.45$\pm$0.15   \\
 \noalign{\smallskip}
12+log(S$^+$/H$^+$)      &     5.77$\pm$0.04	 &   5.50$\pm$0.04 	&  5.75$\pm$0.09    &  5.67$\pm$0.05    &    5.54$\pm$0.09       &    5.41$\pm$0.09  &   5.40$\pm$0.05   \\
log(S$^+$/O$^+$)            &   -2.19$\pm$0.16   &  -2.28$\pm$0.14  & -2.14$\pm$0.33    & -2.17$\pm$0.17    &   -2.20$\pm$0.28       &   -2.29$\pm$0.29  &  -2.25$\pm$0.18  \\ 
log(N$^+$/S$^+$)            &    0.77$\pm$0.09   &    0.83$\pm$0.09  &  0.74$\pm$0.24   &   0.76$\pm$0.10    & {\bf 0.95$\pm$0.19}  &     0.83$\pm$0.20  &   0.79$\pm$0.11  \\
 \noalign{\smallskip}
12+log(Ne$^{++}$/H$^+$) &    7.50$\pm$0.16	 &   7.45$\pm$0.19 	&      \nodata            &  7.50$\pm$0.25    &        \nodata            &         \nodata             &  7.38$\pm$0.21\\
12+log(Ne/H)                   &     7.80$\pm$0.16	 &   7.65$\pm$0.18 	&      \nodata            &  7.73$\pm$0.25    &        \nodata            &          \nodata            &  7.52$\pm$0.21\\
log(Ne/O)                          &    -0.46$\pm$0.23	 &  -0.58$\pm$0.24 	&      \nodata            & -0.51$\pm$0.30    &        \nodata            &         \nodata             & -0.69$\pm$0.27  \\
\noalign{\smallskip}
 12+log(Ar$^{++}$/H$^+$) &     5.90$\pm$0.08	 &   5.92$\pm$0.09 	&  5.81$\pm$0.16    &  5.94$\pm$0.10    &    5.90$\pm$0.13        &    5.90$\pm$0.13   &   5.96$\pm$0.09  \\
 log(Ar$^{++}$/O$^{++}$)  &     -2.06$\pm$0.14 & -2.11$\pm$0.16  &  -2.13$\pm$0.26  &  -2.07$\pm$0.17   &   -2.12$\pm$0.21      &      -2.15$\pm$0.22  &  -2.11$\pm$0.17    \\                    
\noalign{\smallskip}
12+log(He$^+$/H$^+$)     &   10.99$\pm$0.04 &   10.98$\pm$0.03 &    11.03$\pm$0.11   &  10.99$\pm$0.04 &  {\bf 11.09$\pm$0.05}  &   10.97$\pm$0.09    &     10.99$\pm$0.03 \\ 
\hline
\end{tabular}
\end{table*}

 \subsection{Chemical Abundances} \label{abpmas}  
 
Because of the lacking of an individual estimation of the electron temperature, we are not able to map the chemical abundances using the direct method. We will discuss in \S\ref{empcal} the effects of using  empirical calibrations to map the che\-mical abundances of an \HII\ region or a starburst galaxy using 2D spectroscopy. Here we analyze the chemical abundances obtained from the integrated spectra of the areas we described in Fig.~\ref{apertures}.

We used the {\sc iraf} package {\sc nebular}  to derive ionic abundances of \ioni{N}{+},  \ioni{O}{+},  \ioni{O}{2+}, \ioni{S}{+}, \ioni{Ne}{+2}, and \ioni{Ar}{+2} from the intensity of collisionally excited lines. We assumed a two-zone scheme for deriving the ionic abundances, adopting \Te[\ion{O}{iii}] for the high ionization potential ions  \ioni{O}{2+}, \ioni{Ne}{+2}, and \ioni{Ar}{+2}, and \Te[\ion{O}{ii}] for the low ionization potential ions \ioni{O}{+} and  \ioni{S}{+}. The errors in the ionic abundances have been calculated as a quadratic sum of the independent contributions of errors in flux, \Ne\ and \Te.

To derive He$^+$/H$^+$, we used three observed lines of \ion{He}{i} at $\lambda\lambda\lambda$4471,~5876, and 6678, weighted by 1:3:1. Case~B emissivities for the \Te[\ion{O}{iii}] and \Ne\ assumed for each region, were taken from the collision-less (low-density limit) calculations by \citet{Bau05} using an on-line available code\footnote{Available at \tt{http://www.pauKy.edu/$\sim$rporter/j-resolved}}.
The collisional to recombination contribution was estimated from \citet{Kin95}, using the interpolation formula provided by \citet{Por07}. The effective recombination coefficients for H$^+$ were taken
from \citet{SH95}.

We computed the total abundances of O and N, adopting O/H = \ioni{O}{+}/\ioni{H}{+} + \ioni{O}{+2}/\ioni{H}{+} to determine the total oxygen abundance and the standard ionization correction factor (ICF) by \cite{PC69}, N/O = \ioni{N}{+} /\ioni{O}{+}, to compute the total nitrogen abundance. For those regions in which we have a detection of the [\ion{Ne}{iii}] lines, we computed the total neon abundance applying the ICF proposed 
by  \cite{PC69}, Ne/O = \ioni{Ne}{+2} /\ioni{O}{+2}.

We derive a total O abundance of \abox=8.26$\pm$0.09 for all the observed region. The derived N/O and Ne/O ratios, \lno=$-1.42\pm0.12$ and log(Ne/O)$\sim-0.42$, are the typical values expected for a galaxy with the metallicity of IC~10 \citep{IT99, Izotov04, LSE10b}. These results are very similar in the regions AP~1 ([HL90]~111c, the brightest star-forming region), AP~2 (the \Ha\ hole) and AP~3 ([HL90]~111d), but they slightly differ in the AP~4 region, which only considers the emission in those two fibers where the WR stars are detected. We will discuss this intriguing result in Sect~\ref{WRenrich}.

 \subsection{Wolf-Rayet content} 
   
We have clearly detected the blue WR bump in two adjacent fibers (spaxels [11,8] and [11,9]). This position coincides with one of the WR stars catalogued within [HL90]~111c (see Fig.\ref{mapa}, right) in the most recent WR stars catalogue of IC~10 \citep{Crowther03}, [MAC92]~24C. We have also a tentative detection of the nebular \ion{He}{ii} $\lambda$4686 line in spaxel [11,10], which corresponds to the position of the WR [MAC92]~24B. However, we do not observe any WR feature at the position of [MAC92]~24A (spaxel [10,10]) or in the position of RSMV~23 (spaxels [2,14] and [2,15]), although the S/N ratio of this last region is relatively low.

We have added the spectra of the two spaxels where the blue WR bump is observed (AP~4) and carefully analysed the properties of the integrated spectrum. The emission line fluxes are listed in Table~\ref{lines}, and the derived physical and chemical properties of the ionized gas are given in Table~\ref{abundancias}. For comparison, we have also analyzed two nearby fibers, the adjacent position to the detection of the blue WR bump (F90, which corresponds to the spaxel [11,10], where there is a tentative detection of the nebular \ion{He}{ii} $\lambda$4686 line), and the fiber with the maximum of the \Ha\ emission (F92, which corresponds to the spaxel [11,12]).  

In our case, the blue WR bump we detect in AP~4 is essentially constituted by the stellar, broad, \ion{He}{ii} $\lambda$4686 emission feature. Figure~\ref{WRfit} shows a detail of this spectrum in the  4620--4750 \AA\ range, showing a prominent bump. We do not detect the \ion{C}{iii}/\ion{C}{iv} $\lambda$4650 or the \ion{C}{iv} $\lambda$5808 (the red WR bump) and hence we agree with the results found by other authors \citep{Crowther03, Vacca+07} that the observed WR features are originated by a WN star. Both the strength of the broad  \ion{He}{ii} $\lambda$4686 emission and the absence of the \ion{N}{v} $\lambda$4604 emission indicate that this is a late-type WN star (WNL).

We have followed the method described in \cite{LSE10a} to estimate the flux of the blue WR Bump. We fitted a broad and a narrow Gaussian for the stellar and nebular \ion{He}{ii} $\lambda$4686 lines. Figure~\ref{WRfit} shows the result of this fitting, including
the residual spectrum after subtracting the best-fit model to the observed spectrum. The best-fit model values are $FWHM_{\rm He\,II\,broad}$=8.12$\pm$0.76~\AA, $F_{\rm He\,II\,broad}$=(8.76$\pm$0.79)$\times 10^{-16}$ erg\,s$^{-1}$\,cm$^{-2}$~for the broad component and $FWHM_{\rm He\,II\,narrow}$=5.4$\pm$0.9~\AA, $F_{\rm He\,II\,narrow}$=(6.1$\pm$1.0)$\times 10^{-17}$ erg\,s$^{-1}$\,cm$^{-2}$ for the narrow component.

Using the reddening coefficient derived for AP~4 and its relation with the extinction at $\lambda$4686 [assuming the \cite{Cardelli89} extinction law with \mbox{$R_V$=3.1],} \mbox{$A_{\lambda 4686}=2.625\times$\CHb,} and considering the distance to IC~10, we derive a total luminosity of \mbox{$L_{He\,II,broad} = (1.35\pm0.17)\times10^{36}$~erg\,s$^{-1}$} for the broad  \ion{He}{ii} $\lambda$4686 emission. We can use this value to compute the total number of WR stars in the burst. For this, and following \cite{LSE10a}, we consider a metallicity-dependence of the WR luminosities. Using the relation between the luminosity of the \ion{He}{ii}~$\lambda$4686 line and the oxygen abundance (their Eq.~7), we estimate that a single WNL with the metallicity found in IC~10 [which we assume to be \mbox{\abox=8.25],}  has a luminosity of $L_{\rm WNL}$(\ion{He}{ii}~$\lambda$4686)=~$1.27\times10^{36}$~erg\,s$^{-1}$. Hence, the blue WR bump observed in this position of IC~10 is consistent with being produced by a single WNL star and not by four, as \cite{Vacca+07}suggested from their optical/NIR images. This conclusion is essentially the same in the case of considering  the luminosity of a WNL star with a solar metallicity to the broad $\lambda$4686 flux,  \mbox{$L_{\rm WNL}$(\ion{He}{ii}~$\lambda$4686)=~$1.6\times10^{36}$~erg\,s$^{-1}$}  \citep{CH06}, which gives 1.19 WNL stars within this starburst.

    \begin{figure}
   \centering
   \includegraphics[width=0.9\linewidth,angle=90]{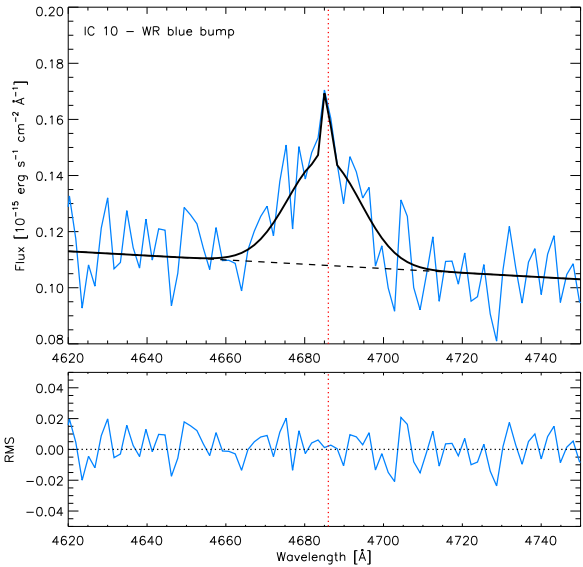} 
   \caption{  \label{WRfit} (Top) Spectrum,  in the 4620--4750 \AA\ range, obtained adding the two fibers (spaxels [11,8] and [11,9]) in which the blue WR bump is observed (AP~4). 
   The red dotted line represents the position of the \ion{He}{ii} $\lambda$4686 emission line, 
   the black dashed line represents the continuum fit. The best fit to the observed spectrum (blue
continuous line) is shown with a black continuous line, and it is composed by the superposition of a broad and a narrow
\ion{He}{ii} $\lambda$4686  lines, following the method explained in L\'opez-S\'anchez \& Esteban (2010a). (Bottom) Residual spectrum after subtracting the best-fit model to the observed spectrum.  }   
  \end{figure}

To estimate the WR/(WR+O) ratio we also follow the prescriptions given by \cite{LSE10a}. We determine this ratio within 
the intense star-forming region. The total number of ionizing photons is \mbox{$Q_{tot}=L_{\rm H\beta}/4.76\times10^{-13}=2.84\times10^{50}$\,s$^{-1}$,} and the \mbox{$\eta$(t)=O7V/O} parameter  
\citep{VC92,V94,SV98} which depends on the age of the burst, is $\eta\sim1$. Hence, applying Eq~10 of \cite{LSE10a}  considering $N_{WNL}$=1 and $N_{WCE}$=0, we estimate a total number of O stars of 26 and a WR/(WR+O) ratio of 0.037. This value agrees with that determined in other WR galaxies with a similar oxygen abundance \citep[see Fig.~5 in][]{LSE10a}.

Fiber~90 (spaxel [11,10]) does not only shows a tentative detection of the nebular \ion{He}{ii} $\lambda$4686 line, but it also shows a clear broad \Ha\ component underlying the narrow, nebular \Ha\ emission. These kinds of features are rather common in extragalactic \HII\ regions \citep[e.g.,][]{Firpo+05, Firpo+10} and starburst galaxies \citep[e.g.,][]{ME97, HomeierGallagher99}. 
Recent detailed analyses of nearby starbursts using IFU spectroscopy [e.g., NGC~1569, \citet{Westmoquette+07a};  M~82, \citet{Westmoquette+07b};  Mkn~996, \citet{James09}] are also revealing the existence of broad components underlying the bright nebular lines that are originated in turbulent movements and shocks of the ionized gas. Our analysis of the emission-line profiles within the center of NGC~5253 \citep{LSEGRPR07} also indicated that they were composed by a narrow and a broad components \citep[see also][]{Monreal-Ibero+10}. Furthermore, the analysis of their chemical abundances revealed that the broad component of the knot~B of NGC~5253 seems to contain the localized N enrichment in this area of the galaxy (see their \S9 and their Table~10). Very different abundance results between individual line components were recently reported in Mkn~996 \citep{James09}. Because of the lacking of spectral resolution, we cannot do this analysis here. However, we have followed
the same procedure used before to fit the nebular and stellar \ion{He}{ii} $\lambda$4686 lines to fit the narrow and broad components of the \Ha\ line in this fiber.
We estimate that the extinction-corrected flux contained in the broad \Ha\ component is $(8.3\pm0.7)\times10^{-15}$\,erg\,s$^{-1}$\,cm$^{-2}$, which is $\sim$11~\% of the flux measured in the narrow line. This value indicates that this component contains a mass of about $\sim$9.5~\Mo. The estimated FWHM of the broad \Ha\ component is $\sim$40~\AA, which indicates a velocity of $\sim$1830~km\,s$^{-1}$. This high value suggests that the flow should be related to a remnant of a recent SN explosion that has still not produced an observable chemical enrichment.


   \subsection{Age of the starburst}
   
Figure~\ref{wha} shows the map of the equivalent width of the \Ha\ emission, \WHa, derived in the intense star-forming region  [HL90]~111 from our PMAS data. As we see, the maximum values of \WHa\ are found at the west of the central star-forming region, which shows \WHa\ between 850 and 1000~\AA. Using the STARBURST~99 stellar population
synthesis model \citep{L99} 
with an instantaneous burst with a Salpeter IMF slope ($\alpha$=2.35) with $M_{up}$=100~\Mo\ and Z=0.008 Z$_{\odot}$ [which corresponds to an oxygen abundance of \abox=8.26\ 
considering a solar oxygen abundance of \abox=8.66 \citep{ASP05} 
we derive an age between 3.3 and 4.2~Myr for the most recent starburst. The eastern areas of this intense star-forming region shows values around \WHa$\sim$600--700~\AA, which are translated to an age of $\sim$4.7~Myr. Similar values are found at the central region of the smaller star-forming region at the east (knot~\emph{d}). The arc-like structure at the south has \WHa$\sim$300-500~\AA.
Although the presence of \Ha\ absorption make underestimate the actual equivalent widths of the emission lines (and hence the age of the last star-formation event is overestimated) we consider that in our case their effect in \WHa\ is very small because of the strength of this emission line. Indeed, the average value of  \Wabs\ derived here (see \S4.1) is only 2.2~\AA. However, as we discussed before, the effect of the \HI\  line absorption should be considered when studying the rest of the Balmer lines and, particularly, to derive a more appropriate value of the reddening coefficient.

  \begin{figure}
\centering
 \includegraphics[width=0.85\linewidth,angle=90]{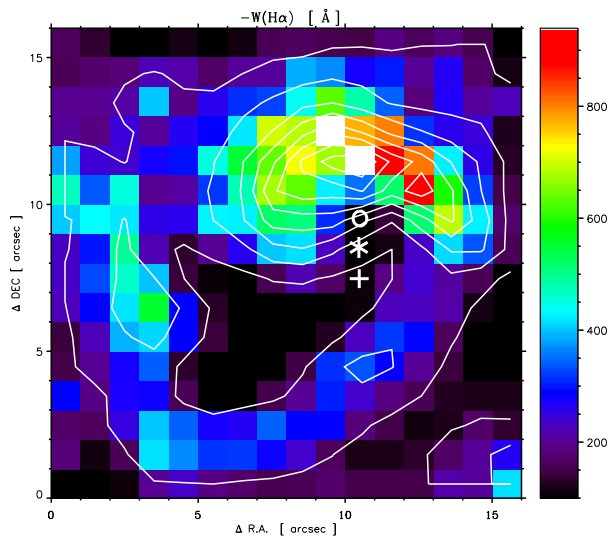}
  \caption{  \label{wha} Map of the equivalent width of the \Ha\ emission, \WHa, derived in the intense star-forming region [HL90]~111 \citep{HL90} from our PMAS data. The contours of the \Ha\ emission are overlaid.  Symbols are the same that in Fig.~\ref{halpha}.}
 \end{figure}

It is interesting to note that the places in which the WR features are detected have relatively low values of \WHa, $\sim$100~\AA, which yields to an age of $\sim$6.2~Myr for the most recent star-formation event. That age is older that that we should expect for a WR star, as the most massive, luminous and hot O stars evolve to the WR phase between 2 and 5~Myr since their birth, spending only some few $10^5$ years before they explode as type Ib/Ic supernovae \citep{MeynetMaeder05}. Hence, 
the star-formation activity may not be strictly instantaneous but it has probably had some continuous rate. Maybe, the detected WR star was formed somewhat later after the majority of the stellar population in the area. Furthermore, as we are considering very small spatial scales (1$\arcsec$--3$\arcsec\sim$3.9--11.6~pc), the WR star may currently be some distance from its birth place\footnote{A star with a proper motion of 20~\kms\ will move $\sim$10~pc in 1~Myr.}.



\section{Discussion} \label{discu}
 \subsection{A possible localized chemical pollution\\ by a Wolf-Rayet star?\label{WRenrich}}

         \begin{figure}
   \includegraphics[width=0.85\linewidth,angle=90]{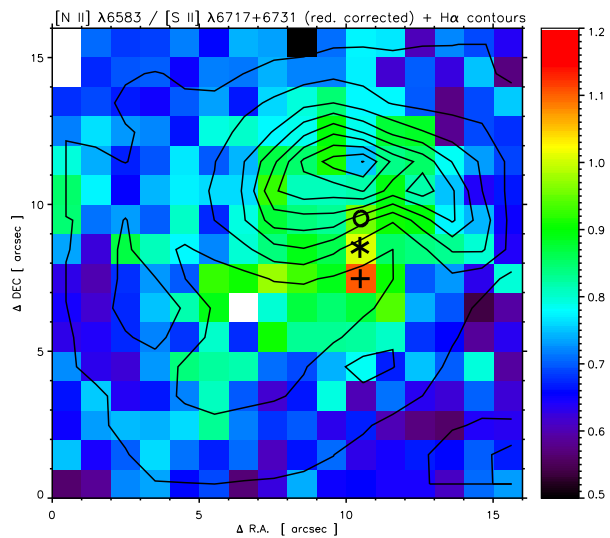} 
   \caption{\label{nitro} 
    [\ion{N}{ii}] $\lambda$6583 / ( [\ion{S}{ii}] $\lambda$6717 + [\ion{S}{ii}] $\lambda$6731) line intensity ratio map derived for the strong star-forming region [HL90]~111 within the starburst galaxy IC~10. The contours of the \Ha\ emission are overlaid. Symbols are the same that in Fig.~\ref{halpha}.}
  \end{figure}	
 
The only confirmed case of localized chemical pollution by WR stars in a starburst region to date is the center of the BCDG NGC~5253 
\citep{WalshRoy89,Kobulnicky97,LSEGRPR07}. This object shows a clear enhancement in the N/H ratio (and also in the He/H ratio) in a small region, which is consistent with the chemical pollution expected by the contribution of just a few (1--2) evolved (WR) massive stars \citep{LSEGRPR07}. 
There are some studies that also suggest that starbursts in which WR stars are detected may show an enhancement in their N/O ratio 
\citep{Pustilnik04, BKD08, James09,LSE10b}.
Recent 2D spectroscopical data \citep{PMV+11} 
also report very high N/O ratios in BCDGs with WR stars. However, in these cases the chemical pollution by the WR stars detected in the central region seems not to be able to explain the large enhancement of nitrogen found throughout the BCDG.
Similarly, \citet*{APMV10greenpeas} used SDSS data to find that, at a given metallicity the so-called \emph{green pea galaxies} display systematically large N/O ratios compared to normal galaxies, but these spectra are not deep enough to detect WR features.

In this case, our 2D spectroscopic PMAS maps of the strong star-forming region  [HL90]~111 within IC~10 suggests the detection of a localized high N/O ratio at the positions where the WR features are detected (AP~4). Figure~\ref{nitro} shows the  [\ion{N}{ii}] $\lambda$6583 / ( [\ion{S}{ii}] $\lambda$6717 + [\ion{S}{ii}] $\lambda$6731) 
line intensity ratio map. Because of the faintness of the [\ion{O}{ii}] $\lambda$3727 doublet, 
the [\ion{N}{ii}] $\lambda$6583 /  [\ion{O}{ii}] $\lambda$3727 line intensity ratio map shows a very low S/N in a large number of spaxels where the surface brightness of the nebular gas is low, and hence it is useless for our purposes.
Considering only the spaxels with higher S/N ratio, the average value of the  [\ion{N}{ii}]~$\lambda$6583~/~[\ion{O}{ii}]~$\lambda$3727  ratio  is $\sim$0.07.  Spaxel [11,8], labelled with a cross in all maps and where some \ion{He}{ii}~$\lambda$4686 emission is observed, has a very uncertain detection of the [\ion{O}{ii}]~$\lambda$3727 doublet. However, spaxel [11,9] (labelled with a star and where the highest  \ion{He}{ii}~$\lambda$4686  emission is observed) seems to have a slightly higher  [\ion{N}{ii}]~$\lambda$6583~/~[\ion{O}{ii}]~$\lambda$3727  ratio, $\sim$0.10, but because of the high uncertainty this value is not significant. On the other hand, the \mbox{[\ion{N}{ii}]~$\lambda$6583 / ( [\ion{S}{ii}] $\lambda$6717 + [\ion{S}{ii}] $\lambda$6731)} ratio, which is well measured in almost all spaxels, indicates a rather clear increase of its average value  ($\sim$0.85) precisely at the two spaxels where WR features are detected. Indeed, spaxels [11,8] and [11,9] show a  [\ion{N}{ii}]~$\lambda$6583~/~([\ion{S}{ii}]~$\lambda$6717+[\ion{S}{ii}]~$\lambda$6731) ratio of 
 1.11 and 1.01, respectively, with a typical uncertainty of $\pm0.12$. 
 
 It is interesting to note that, the \ion{He}{i} $\lambda$5875/\Ha\ ratio map (top right panel in Fig.~\ref{mapas}) also shows an enhancement of the emission at the same position. The average  \ion{He}{i} $\lambda$5875/\Ha\ ratio is $\sim$0.046 
 but spaxels [11,8] and [11,9] have a ratio of 0.071 and 0.058,  
 respectively, with an error of only $\pm$0.007. 
It is possible that we are detecting at this position a localized enhancement of both nitrogen and helium, which, if true, is very probably connected with the detection of WR features in those same spaxels.
 
 We further investigate this very intriguing result via the careful analysis of the chemical properties of the integrated spectrum obtained adding both spaxels (AP~4). As control, we also study the adjacent spaxel [11,10] (Fiber~90), which has enough S/N ratio to perform such analysis. Table~\ref{abundancias} compiles all the chemical abundances and ratios derived for these two regions. As we see 
 AP~4 possesses, systematically, higher N/O, N$^+$/S$^+$ and He$^+$/H$^+$ ratios ($-1.26$, 0.95 and 11.09, respectively) than those values derived for \emph{any} other aperture or spaxel, which have average values of $-1.44$, $0.79$, and 10.99, respectively. We also notice that the S$^+$/O$^+$ (two elements of similar nucleosynthetic  origin) ratio remains practically the same for all regions including AP~4, showing a average value of $-2.22$. 
 Hence, although we cannot definitively confirm that there is a localized chemical pollution in nitrogen in those two spaxels 
 because of the relatively high uncertainties, all evidence points towards it.
 We must emphasize that the helium enrichment is indeed confirmed within the errors, because the average value of 12+log(He$^+$/H$^+$) is $10.99\pm0.03$ while that measured in AP~4 is $11.09\pm0.05$.

 With the data collected in tables~\ref{lines} and \ref{abundancias} we can make a rough estimation of the mass of newly created He and N (stellar yield) necessary to produce the observed overabundances in region AP~4. These yields have been computed considering the He and N abundances of region AP~1 as reference. Therefore, we have assumed that the initial abundances of the ionized gas at AP~4 are those measured at AP~1 and that AP~4 has suffered a localized and very recent increment of the He/H and N/H ratios. Finally, the empirical stellar yields have been obtained multiplying those He/H and N/H increments by the mass of ionized gas derived for AP~4. The computed stellar yields are 28$\pm$16~$M_{\odot}$ and 0.013$\pm$0.013$~M_{\odot}$ for He and N, respectively. These numbers can be compared with those given in table~12 of \citet{LSEGRPR07}, that includes the yields determined for regions A and B of NGC~5253 as well as theoretical ones from stellar evolution models and empirical determinations for ring nebulae around Galactic WR stars. The yields found in [HL90]~111c (although uncertain) are roughly consistent with the scenario of  chemical pollution by the ejecta of a very small number of WR stars. The amount of He appears to be rather high, in fact the ratio of He and N stellar yields is about 2150, ten times higher than in the objects included in the aforementioned table of \citet{LSEGRPR07}.

We must say that it is not the first time that a region with a probably high N/O is reported in IC~10. \citet{MagriniGoncalves09}
used Gemini North Multi-Object Spectrograph to map the \HII\ regions and Planetary Nebulae (PN) in the central 5.5'$\times$5.5' of IC~10. Their data suggested that the \HII\ region [HL90]~120, 
 with an oxygen abundance of \abox$\sim$8.18, has a N/O ratio of $\sim-$0.48 dex, but the other \HII\ regions (which have an oxygen abundance between 8.0 and 8.7) show N/O ratios of $\sim-$1.25. However, their data have high uncertainties (the majority of N/O values agree within the errors, see their Fig.~2), and hence this result should be also confirmed. Furthermore, all the N/O ratios these authors derived for their analyzed \HII\ regions within IC~10 seem to be higher than those expected for their oxygen abundance.  

As \citet{MagriniGoncalves09} pointed out, the \HII\ region [HL90]~120 contains four known WR stars within 20$\arcsec$ from its center, and hence the origin of the tentative high N/O ratio may be some of these WR stars. In our case, we notice that our possible detection of a high N/O and He/H ratios are found \emph{exactly} at the same position where the WR features are located, indicating that, if real, this chemical pollution is indeed very localized ($\sim2\arcsec\sim7.8$~pc) and therefore it should be very difficult to detect in fainter and distant galaxies.

Some studies affirm that the more or less low and constant values of the N/O ratio reported in BCGs is a consequence of the fact that anything ejected by massive stars (whether oxygen by type~II supernovae or nitrogen in the winds of WR stars) is too hot to be immediately incorporated to the ISM, and it needs to cool before mixing with the existing nebular gas and appears in the optical spectrum 
\citep*[e.g.,][]{KS96,KS97,KS98, Kobulnicky97,Henry00,PTV03,vZH06,MagriniGoncalves09, Lagos09}. 
However, as previously suggested by \citet{Pustilnik04} and \citet{LSEGRPR07}, the presence of localized chemical enrichment in very young
starbursts indicates that the timescale of the process should be very short. This agrees with the enrichment pattern observed in Galactic WR ring nebulae 
\citep{EstebanVSC92}. 
The statistical analysis performed by \citet*{BKD08} comparing WR and non-WR galaxy data from SDSS observations also yields to a similar conclusion. The recent,  more detailed, analysis of a sample of WR galaxies performed by \citet{LSE10b} also reported high N/O ratios in 6 WR galaxies. These authors  indicated that the high N/O values should dilute in a quick period because of both the decreasing of the star-formation activity and the releasing of new, fresh material which is dispersed and mixed with the existing gas of the galaxy. Furthermore, the effects of the WR stars into the ISM are also diluted because of aperture effects 
\citep{LSE09}.
Only very careful analyses of such systems \citep[e.g.,][]{LSE10b}
will help to solve the puzzle of the high N/O ratios sometimes found in starburst galaxies.

 \subsection{Use of empirical calibrations in 2D spectroscopy data\label{empcal}}

  \begin{figure*} 
   \centering
    \begin{tabular}{cc}
 \includegraphics[width=0.4\linewidth,angle=90]{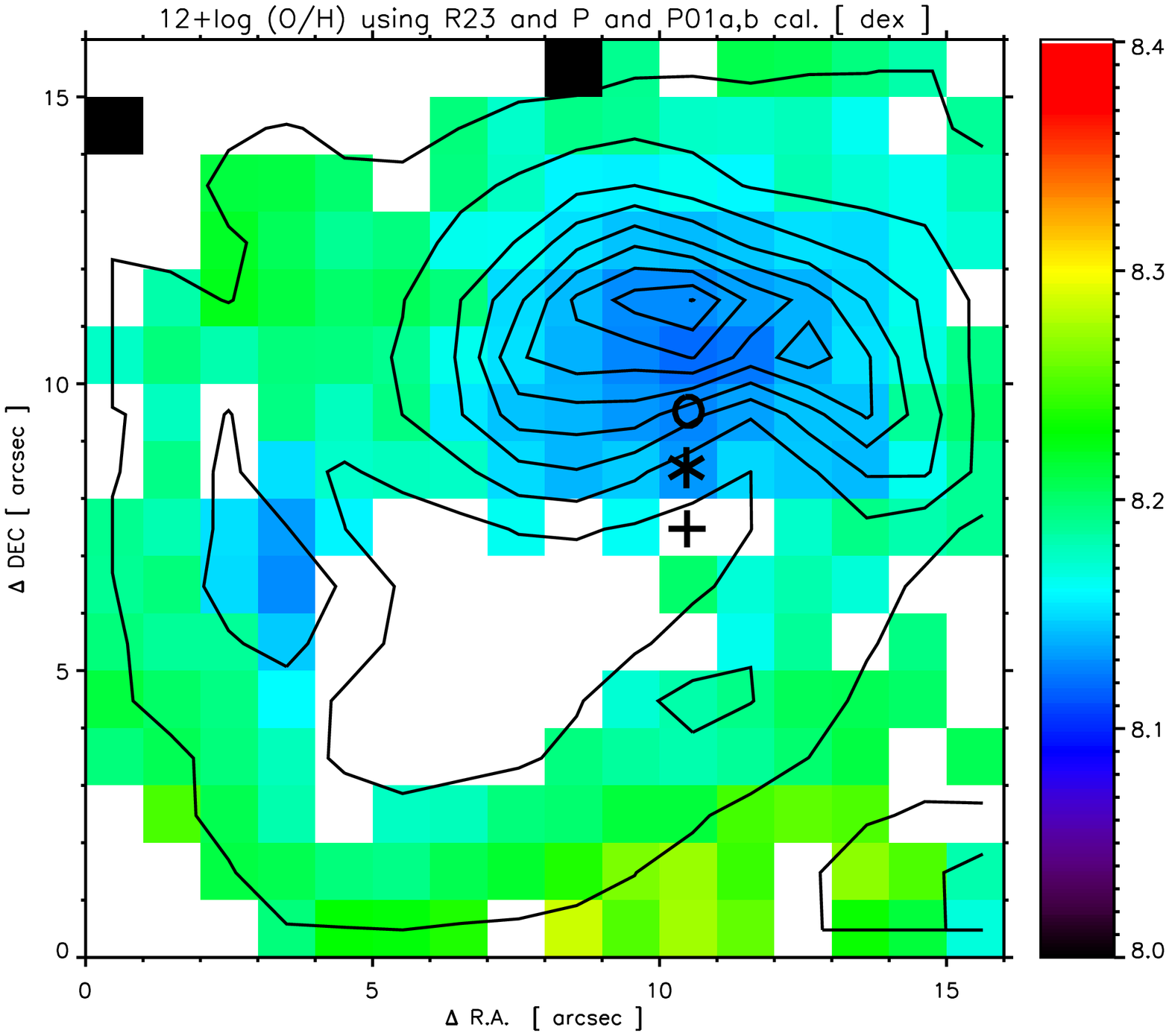} &
  \includegraphics[width=0.4\linewidth,angle=90]{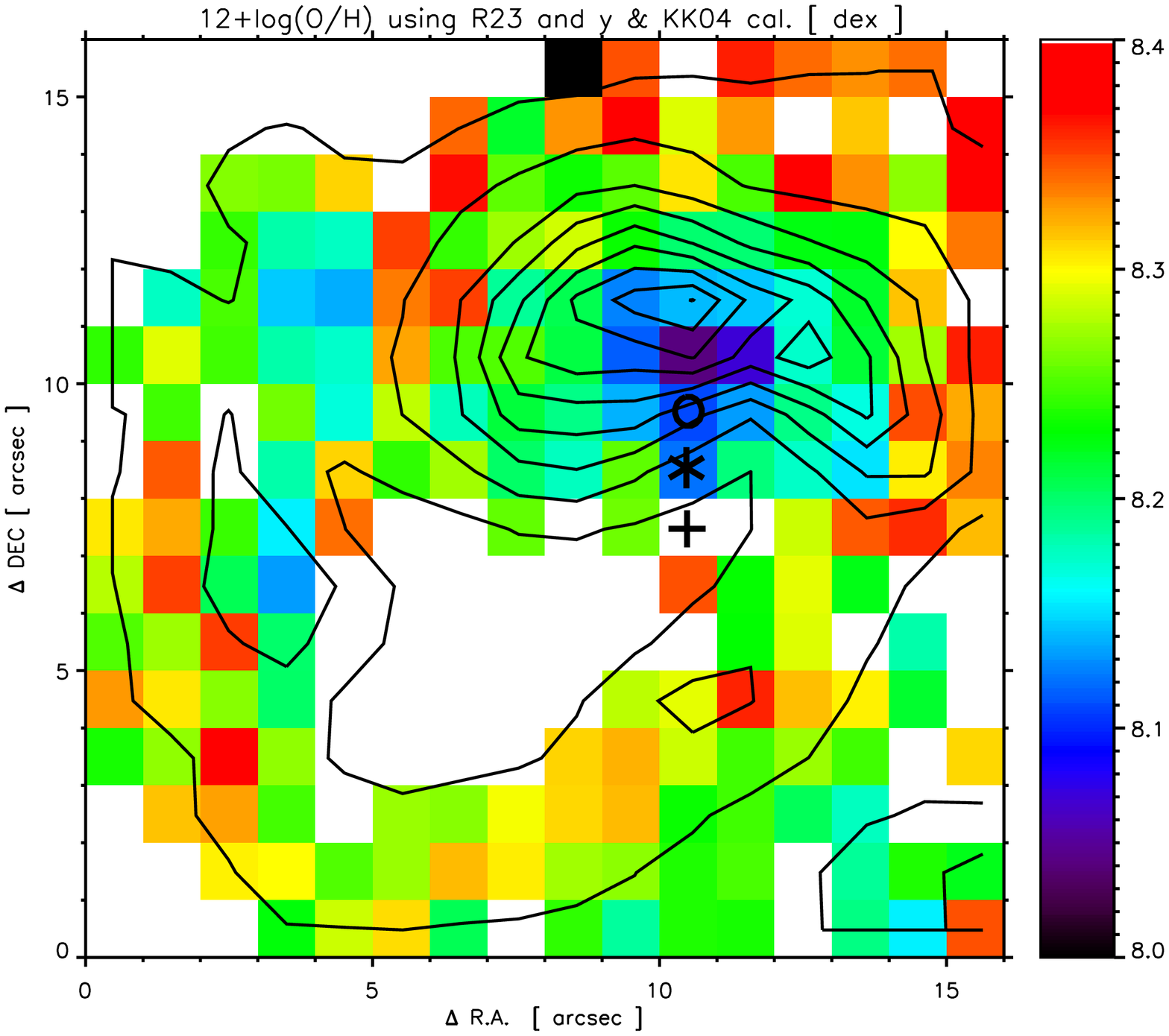} \\
 \includegraphics[width=0.4\linewidth,angle=90]{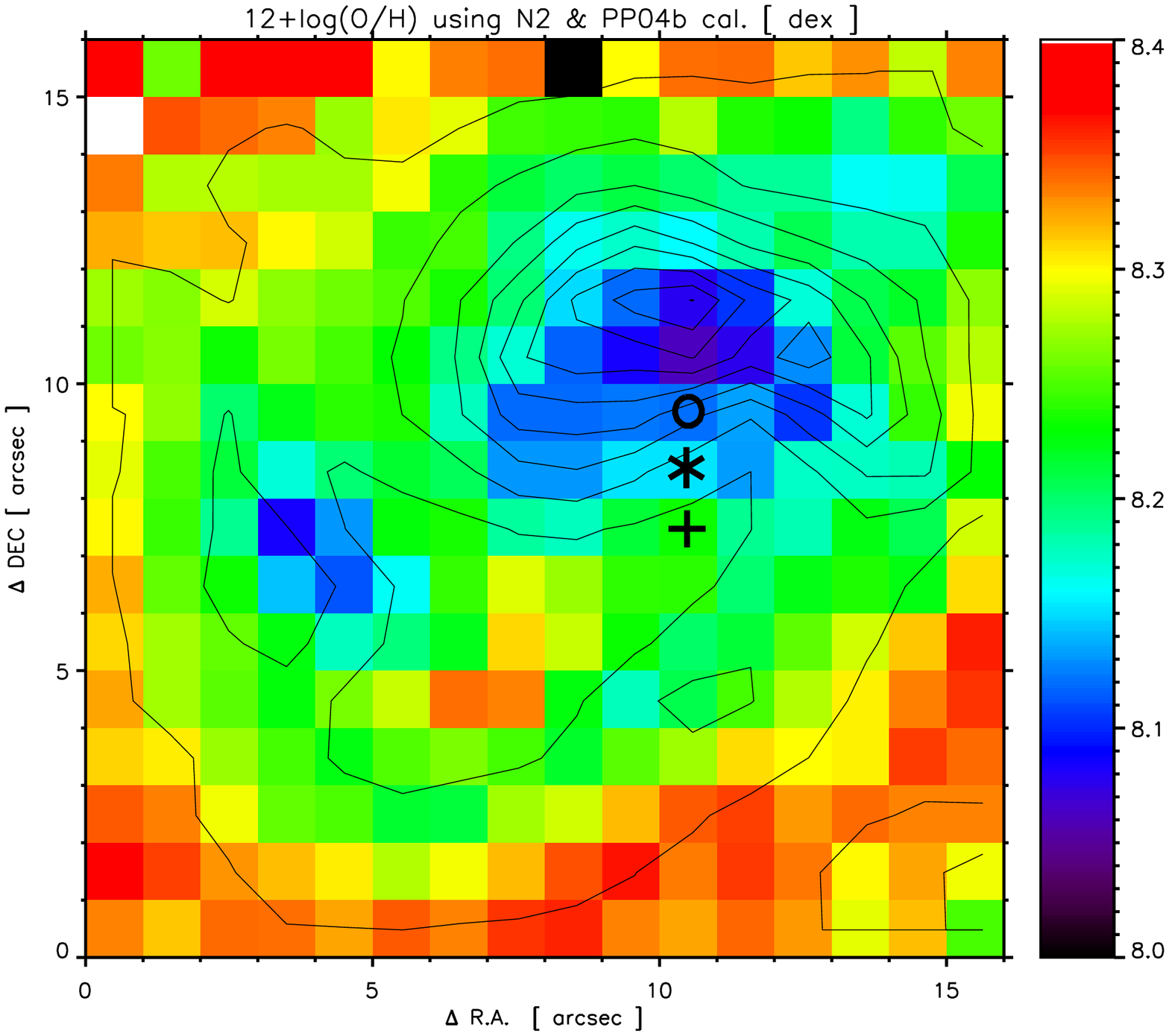} &
 \includegraphics[width=0.4\linewidth,angle=90]{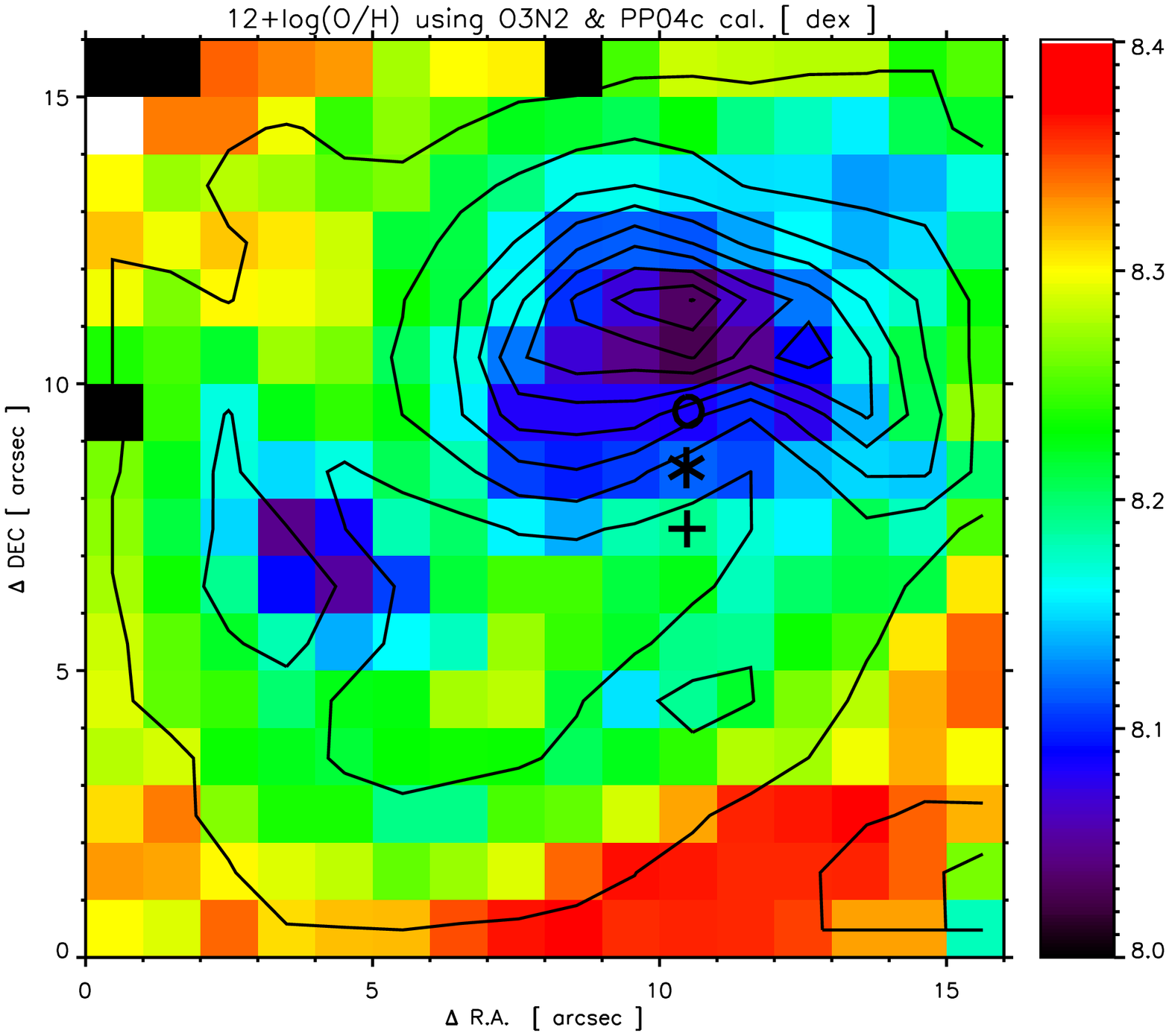} \\
 \end{tabular}
   \caption{   \label{empirica} Maps of the oxygen abundance derived by empirical calibrations  with H$\alpha$ contours overlaid into the IC~10 PMAS field
            of: (\emph{tot left}) Pilyugin calibration [average value between  \citet{P01a,P01b}] using the $R_{23}$ and $P$ parameters,
            (\emph{top right}) \citet{KK04}  calibration using the $R_{23}$ and the $y$ parameters,
                       (\emph{bottom left})  cubic calibration between the $N_2$ parameter and the oxygen abundance derived by \citet{PP04}, and     
                                 (\emph{bottom right})   cubic calibration between the $O_3N_2$ parameter and the oxygen abundance derived by \citet{PP04}.  
                                 Symbols are the same that in Fig.~\ref{halpha}.}
 \end{figure*}
  
\begin{table*} 
\caption{\label{abunempiricas} Oxygen abundances, in the form \abox, of the regions analyzed in IC~10 using our PMAS data determined using empirical calibrations.}
\begin{tabular}{l   c c cc cc  }
\hline
 Method     & Parameters  & 	  [HL90] 111       &  AP1 (\emph{c})     &  AP2 (\emph{hole})       &    AP3 (\emph{d})      &     AP4 (WR 24)     \\
\hline
  
 Direct       & \Te                         &     8.26$\pm$0.09	 &   8.22$\pm$0.08 	&  8.22$\pm$0.17    &  8.23$\pm$0.09    &    8.20$\pm$0.12       \\
\noalign{\smallskip}
P01           & $R_{23}$, $P$       &     8.15    &   8.12      & 8.10   & 8.13   &  8.11 \\
PT05       & $R_{23}$, $P$       &     8.13   &   8.10  &   8.03 &   8.13  &   8.07 \\
KK04         & $R_{23}$, $y$       &     8.25    &   8.17     &  8.00   & 8.22   & 8.10 \\
PP04b       & $N_2$                   &     8.22     & 8.13    & 8.08    & 8.18   &  8.15 \\
PP04c       & $O_3N_2$             &     8.20    &  8.11   & 8.07    & 8.15  &   8.12 \\

\hline
\end{tabular}
\end{table*}

We also study the results provided by the so-called empi-rical calibrations of the oxygen abundance [see \citet{KE08} and \citet{LSE10b} for recent reviews] when analyzing data provided by 2D optical spectroscopy, which in many occasions have a good determination of all important nebular lines involved in the most common empirical calibrations (i.e., [\ion{O}{ii}]~$\lambda$3727, \Hb, [\ion{O}{iii}]~$\lambda\lambda$4959,5007, \Ha,  [\ion{N}{ii}]~$\lambda$6583). Ratios between these bright emission lines define the parameters $R_{23}$, $P$, $y$, $N_2$, and $O_3N_2$ [see definitions in Appendix~A of \citet{LSE10b}].
Empirical calibrations are very commonly used to get a rough estimation of the metallicity of star-forming galaxies when the electron temperature of the ionized gas cannot be determined (i.e., there are no detections of the faint auroral lines).

Although some precautions must be taken into account when analyzing star-forming galaxies with intermediate metallicities, 7.9$\lesssim$\abox$\lesssim$8.4, 
\citep{Yin+07}, it seems that the combination of the $R_{23}$ and an excitation parameter ($P$ or $y$) have the necessary information for the determination of accurate abundances in extragalactic \HII\ regions.
The comparison between the oxygen abundances provided by empirical calibrations and those determined via the direct method performed by 
\citet{LSE10b} concluded that the Pilyugin-method \citep{P01a,P01b,PT05} 
is nowadays the best suitable empirical calibration to derive the oxygen abundance of star-forming galaxies. 
However, the results provided by empirical calibrations based on photoionization models 
\citep{McGaugh91,KD02,KK04}
are systematically 0.2--0.3 dex higher than those found following the direct method. 
This result has been later confirmed by \citet{Moustakas+10},
who found differences of $\sim$0.6 dex between both kinds of method.
On the other hand, \citet{Yin+07} indicated that the  $N_2$  and the $O_3N_2$ indices are useful for calibrating metallicities of galaxies with \abox$<$8.5. 
\citet{LSE10b} concluded that the relations between these two parameter and the oxygen abundance provided by \citet{PP04} give acceptable results for objects with \abox$>$8.0. \citet{PerezMonteroContini09} established that empirical calibrations using the $N_2$ parameter are not valid for objects with \abox$\lesssim $8.0. However, the effects of such empirical calibrations using 2D spectroscopic data have not been explored so far. 

Figure~\ref{empirica} shows the oxygen abundance maps of the observed star-forming region [HL90]~111 within IC~10 derived from four of the most common empirical calibrations:  \citet{P01a,P01b},
which involve the $R_{23}$ and the $P$ parameters; \citet{KK04},
which is based on the photoionization models presented by \citet{KD02}
and considers the $R_{23}$ and the $y$ parameters to iteratively derive the oxygen abundance and the ionization parameter $q$; the bicubic relation between the oxygen abundance and the $N_2$ parameter derived by 
\citet{PP04};
and the linear fit between the oxygen abundance and the $O_3N_2$ parameter also determined by \citet{PP04}.
As the oxygen abundance determined using the direct method in IC~10 is found between the low-metallicity [\abox$\leq$8.1] and the high-metallicity [\abox$\geq$8.4] branches, we assumed the average value between the \citet{P01a,P01b} calibrations, but always the metallicity derived for the low branch following the \citet{KK04} calibration. The four panels shown in Figure~\ref{empirica} have the same scale in the oxygen abundance, between 8.0 and 8.4.

The oxygen abundance map provided by the \citet{P01a,P01b} calibration (top left panel in Fig.~\ref{empirica}) shows a more or less constant value, although it clearly has lower values in the high-ionization regions [\abox$\sim$8.13] that in the low-ionization regions [\abox$\sim$8.21], but both agree within the errors ($\pm$0.10 dex) and, although slightly lower, with the value determined using the direct method [\abox=8.26$\pm$0.09]. If using the \citet{PT05} calibration, the resulting metallicity map has a similar structure to that shown in top left panel of Fig.~\ref{empirica}, but the derived values are systematically $\sim$0.05--0.10~dex lower than those determined following the \citet{P01a,P01b} method. 
In any case, as we will discuss below it is evident that there is some effect of the ionization degree on the determination of the oxygen abundance and that this calibration seems to be not suitable for looking for chemical differences smaller than $\sim$0.10--0.15~dex using 2D spectroscopic maps.

The situation is somewhat similar when using the \citet{KK04} calibration (top right panel in Fig.~\ref{empirica}). As we see, this calibration provides oxygen abundances between 8.04 and 8.38, with the lower values located again in the high-ionization regions. Spaxels which have higher oxygen abundances following this method are typically located in the external areas of the star-forming regions, where the ionization degree is lower and the uncertainty in the [\ion{O}{ii}]~$\lambda$3727 flux is higher. In any case, we notice that the  results derived from the \citet{KK04} calibration have a higher dispersion that those obtained following the Pilyugin method, and therefore this calibration should be used with caution when analyzing the chemical abundances of 2D spectroscopic maps.

    \begin{figure}
   \centering
   \includegraphics[width=0.82\linewidth,angle=90]{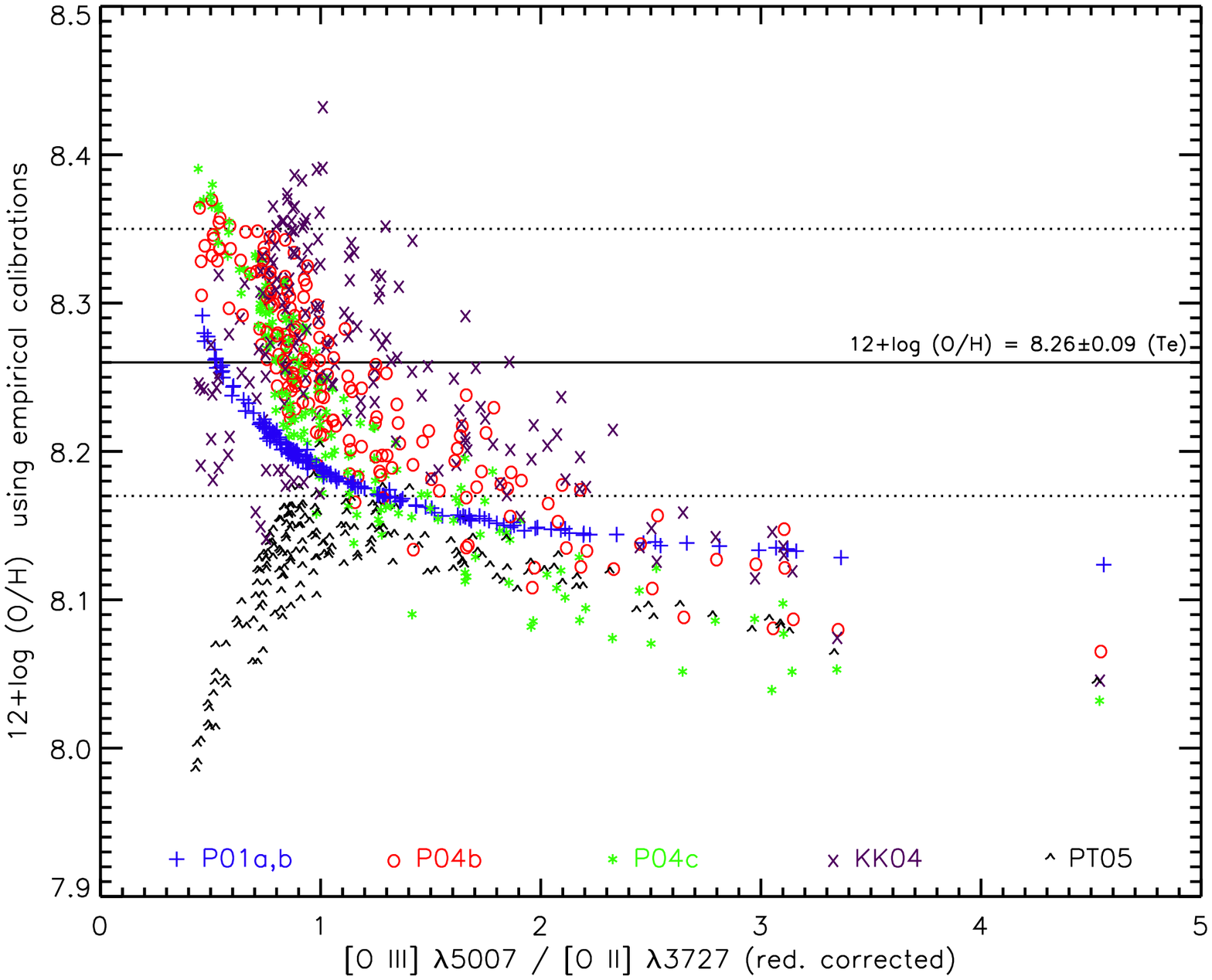} 
   \caption{  \label{empiricao3o2} Relation between the ionization degree, given by the reddening-corrected [\ion{O}{iii}]~$\lambda$5007~/~[\ion{O}{ii}]~$\lambda$3727 ratio ($x-axis$), and the oxygen abundance ($y-axis$) derived from empirical calibrations for the spaxels that possess a good measurement of the [\ion{O}{ii}]~$\lambda$3727 emission line. Different symbols and colors indicate different empirical calibrations: \citet{P01a,P01b} (blue crosses) and \citet{PT05} (black $\wedge$);
   \citet{KK04} using $R_{23}$ and $y$ parameters (purple x); and \citet{PP04} using the $N_2$ (red circles) and the $O_3N_2$ (green stars) parameters. The oxygen abundance determined in the region [HL90]~111 using the direct method, \abox=8.26, is indicated with a black continuous line. The two black discon-tinuous lines shows the uncertainty, $\pm$0.09, of this value.}   
  \end{figure}

The ionization structure of the giant \HII\ regions we are observing with our PMAS data is indeed playing an important role in the determination of the oxygen abundances using empirical calibrations. That is very evident when using the $N_2$ and $O_3N_2$ parameters, which strongly depend on the ionization degree of the gas.  The oxygen abundance map obtained using the \citet{PP04} calibrations involving the $N_2$ (bottom left panel in Fig.~\ref{empirica}) and the $O_3N_2$  (bottom right panel in Fig.~\ref{empirica}) parameters directly reflect the ionization structure (see Fig.~\ref{o3o2}), having lower oxygen abundances [\abox$\sim$8.10 and 8.05, respectively] in the high-ionization region and higher oxygen abundances [\abox$\sim$8.30] in the low-ionization areas. 

The correlation between the ionization degree and the results provided by the empirical calibrations is shown in Fig.~\ref{empiricao3o2}, that plots the reddening-corrected [\ion{O}{iii}]~$\lambda$5007~/~[\ion{O}{ii}]~$\lambda$3727 ratio versus the oxygen abundance derived from empirical calibrations for the spaxels that possess an useful measurement of the [\ion{O}{ii}]~$\lambda$3727 emission line. From this figure is evident that empirical calibrations give lower oxygen abundances in regions with higher ionization degree, with a monotonic increasing of the oxygen abundance with decreasing ionization degree in all cases except for the \citet{PT05} calibration (which changes this tendency for [\ion{O}{iii}]~$\lambda$5007~/~[\ion{O}{ii}]~$\lambda$3727 $\lesssim$1.0).
Figure~\ref{empiricao3o2} also shows that the results provided by the \citet{KK04} method have a higher dispersion than those obtained using other empirical calibrations, and that they tend to be between 0.1 and 0.2 dex higher than the others.

However, when we integrate the individual spectra of an \HII\ region, the result provided by the empirical calibra\-tions are closer to those derived from the \Te\ method. Table~\ref{abunempiricas} lists all oxygen abundances determined using empirical calibrations in all the regions analyzed here, including all the observed area. As we see, \citet{P01a,P01b} calibration provides a constant value of \abox$\sim$8.12, that is $\sim$0.14 dex lower than the oxygen abundance determined by the direct method. \citet{PT05} calibration provides somewhat lower values, \abox$\sim$8.10.
 The \citet{KK04}  method gives a relatively good value for the bright \HII\ regions [\abox$\sim$8.20], but the values obtained for the \emph{hole} and AP~4 (the two spaxels where the blue WR bump is detected) are much lower. 
 When integrating the flux of all the \HII\ region, the \citet{PP04} calibrations using the $N_2$ and $O_3N_2$ indices also provide oxygen abundances that agree within the errors with those determined using the direct method.

Beside the ionization structure of the \HII\ regions, other biases  (such as the hardness of the ionizing radiation field, the structural parameters of the \HII\ regions or the star formation history of the host galaxy) may also affect to the determination of the chemical abundances using empirical calibrations \citep{Stasinska10}.
Hence, we conclude that the analysis of star-forming galaxies (and strong starbursting, Wolf-Rayet galaxies in particular) using 2D spectroscopy needs observations deep enough to detect (at least in the brightest \HII\ regions) the faint auroral lines, if a good determination of the chemical abundances and physical conditions of the ionized gas is wanted to be achieved \citep[e.g.,][]{James09,James10}. The deep observations will also provide a better estimation of the [\ion{O}{ii}]~$\lambda$3727 flux, which is essential to precise the total oxygen abundance. In the case this doublet cannot be observed, deep observations at $\lambda$7330 are needed to get a good determination of the [\ion{O}{ii}]~$\lambda\lambda$7318,7330 emission lines, which typically are 20--30 times fainter than [\ion{O}{ii}]~$\lambda$3727, but this spectral range is very affected by telluric lines. Oxygen abundance maps obtained using empirical calibrations may show features that are not related with the actual metallicity distribution of a star-forming galaxy but with the ionization structure within its giant \HII\ regions.

\section{Conclusions} \label{conclu}

In this paper, we present results from integral field spectroscopy of the intense star-forming region [HL90]~111 within the starburst galaxy IC~10. This study is based on PMAS data, which cover a wavelength range of 3500--7150~\AA, with an effective spectral resolution of 3.6~\AA, and map an area $16\arcsec\times16\arcsec$ with a spatial sampling of $1\arcsec\times1\arcsec$, that corresponds to 3.88 pc$^2$ at the distance to IC~10.
We obtain maps of different emission lines and analyzed the extinction, physical conditions, nature of the ionization, and chemical abundances of the ionized gas, as well determined locally the age of the last star-formation event. 

By defining several apertures, we study the main integrated properties of some regions within [HL90]~111. For the whole FoV, we derive an extinction in the $V$-band of $A_V=3.05\pm0.11$~mag, a total mass of ionized gas of $(1.76\pm0.10)\times~10^{5}$ \Mo, a star-formation rate density of \mbox{$1.64\times0.09$~\Moy\,kpc$^{-2}$,}  an oxygen abundance of \abox=8.26$\pm$0.09 and a nitrogen to oxygen ratio of \lno=$-1.42\pm0.12$. The age of the most recent star-formation event is $\sim$3.3~Myr.

One of our main aims is  to precise the places where Wolf-Rayet stars are located via the detection of the blue and/or red WR bumps.
Only two spaxels show an unambiguous detection of the broad \ion{He}{ii}~$\lambda$4686 emission line. This feature seems to be produced by a single WNL star, its position coincides with [MAC92]~24B.
We report a possible N and He enrichment at the same location where the WR feature is found. Moreover, the enrichment pattern is roughly consistent with that expected for the pollution of the ejecta of a single or a very small number of WR stars. This result suggests that the fresh material released by massive stars is quickly incorporated into the surrounding ISM. 
Furthermore, this chemical pollution is very localized ($\sim2\arcsec\sim7.8$~pc) and it should be difficult to detect in star-forming galaxies beyond the Local Volume.

The reddening coefficients derived from the \Ha/\Hb\ and the \Hg/\Hb\ ratios have a better agreement when the absorption underlying the \HI\ Balmer lines is considered.
We find a correlation between the electron density of the ionized gas and the reddening coefficient, indicating that denser regions have higher extinction. 
Furthermore, the low reddening coefficients computed around the spaxels where the WR features are detected suggest that the gas surrounding the massive stars has been evacuated or partially destroyed. In the spaxel just to the north of the WR stars location we also detect a broad \Ha\ component underlying the nebular \Ha\ emission that may be related to a remnant of a recent SN explosion.

We discuss the use of the most-common empirical cali-brations to estimate the oxygen abundances of the ionized gas in nearby galaxies from 2D spectroscopic data. The ionization degree of the gas plays an important role when \mbox{applying} these empirical methods, as they tend to give lower oxygen abundances with increasing ionization degree. We finally stress the importance of the detection of the auroral lines to derive the electron temperature of the ionized gas and obtain a precise map of the chemical abundances in extragalactic \HII\ regions.

\section*{Acknowledgments}

We thank the referee, Mark Westmoquette, for his very encouraging and detailed reviewing, that improved the quality of this paper.
We are indebted to Jorge Garc\'{\i}a-Rojas for very fruitful discussions and comments to this study and for his help to derive the ionic abundances of helium.
We are also grateful to Enrique P\'erez-Montero for his careful reading of the manuscript and his very interesting comments. 
The authors thank Sebasti\'an S\'anchez for his support and help during our observations at the 3.5m telescope at Calar Alto Observatory. 
\'A. R. L-S. thanks the people at the Instituto de Astrof\'{\i}sica de Canarias for their friendship during his stay in Tenerife in August and September 2010.
This research has made extensive use of the
SAO/NASA Astrophysics Data System Bibliographic Services (ADS).


\label{lastpage}

\end{document}